\documentclass[prb,twocolumn,showpacs,floatfix,superscriptaddress,amsfonts]{revtex4-1}
\usepackage{graphicx,graphics,color,epsfig}% Include figure files
\usepackage{bm}
\usepackage{amsmath}
\usepackage{amssymb}

\begin{document}
\preprint{}
\title{Mott Transition in Multi-Orbital Models for Iron Pnictides}

\author{Rong Yu}
\affiliation{Department of Physics \& Astronomy, Rice University, Houston, Texas 77005, USA}
\author{Qimiao Si}
\affiliation{Department of Physics \& Astronomy, Rice University, Houston, Texas 77005, USA}

%\date{\today}
\begin{abstract}
The bad-metal behavior of the iron pnictides has motivated a theoretical
description in terms of a proximity to Mott localization. Since the parent
compounds of the iron pnictides contain an even number of 3d-electrons
per Fe, it is important to determine whether a Mott transition
robustly exists and clarify the nature of the possible Mott insulating phases.
We address these issues in a minimal two-orbital model and a more realistic
four-orbital model for the parent iron pnictides using a slave-spin
approach.
In the two-orbital model with two electrons per Fe, we identify
a single transition from
%metal to Mott insulator,
a metal to a Mott insulator,
showing that this transition must exist as a result of orbital degeneracy.
%as the consequence of orbital degeneracy.
%The critical coupling, $U_c$, is greatly reduced by the
%Hund's coupling.
Depending on the ratio between the inter- and intra-orbital Coulomb repulsions,
the insulating state can be either a high-spin Mott insulator or a low-spin orbital-Mott insulator.
In the four-orbital model with four electrons per Fe, we find a rich phase diagram for the metal-to-insulator transition. At strong Hund's couplings, a localization transition to a high-spin Mott insulator always occurs. At zero and weak Hund's couplings, on the other hand, we find a transition to an intermediate spin insulating state. This transition can be viewed as an orbitally selective metal-to-insulator transition: the transition to a Mott insulator in the $xz$ and $yz$ orbitals takes place at the same critical coupling as the transition to either a band insulator at zero Hund's coupling or an orbitally polarized insulator at weak but finite Hund's coupling in the $xy$ and $x^2-y^2$ orbitals.
The implications of our model studies for the physics of iron pnictides and iron chalcogenides
are discussed.
\end{abstract}
\pacs{71.10.Hf, 71.27.+a, 71.55.-i, 75.20.Hr}
\maketitle

\section{Introduction}
The microscopic physics of the iron pnictides and related
high-$T_c$ superconductors~\cite{Kamihara_FeAs,Zhao_CPL08}
is a subject of extensive studies. The parent systems are
antiferromagnetically ordered \cite{delaCruz},
implying that Coulomb interactions must play some role.
The metallic nature of these systems gives rise to
%the
a
tendency
to treat the interactions perturbatively. However,
various considerations have
%advanced
led to
the notion that
the parent iron pnictides and iron chalcogenides
are on the verge of a Mott localization transition.
%Such
These
considerations have been based on the observed bad-metal
properties~\cite{Si-Abrahams-prl08,Si-Abrahams-Dai-Zhu},
first-principles calculations~\cite{Haule-Kotliar-prl08,Laad08},
and related
analyses ~\cite{Daghofer08,CFang08,CXu08,IshidaLiebsch,Leeetal09}.
%For the bad-metal behavior,
%there are several aspects that
The bad-metal properties
are characteristic of metallic systems in proximity
to a Mott localization. The electrical resistivity
at room temperature corresponds to a short mean free path,
on the order of the average inter-electron spacing.
The Drude weight seen in the room temperature
optical conductivity
is considerably suppressed compared to its non-interacting value
\cite{Qazilbash,Hu08,Si_natphys}. Finally, relatively small changes
of temperature induce transfers of the optical spectral weight
extending to the eV range \cite{Hu08,Yang08,Boris09}.
The lack of observation\cite{Yang09}
of any pronounced
incoherent peaks in the high-energy electron spectrum
has raised some questions about the incipient Mott
picture, but recent microscopic calculations \cite{Kutepov10}
have suggested that this
%is due to
arises from a large damping
%effects
whose effect is
enhanced
by the multi-orbital nature of the system.

There is some evidence for the incipient Mott picture from the
magnetic sector as well. High-energy spin-wave-like excitations
have been seen at the zone boundaries \cite{Zhao}.
In addition, the total spin spectral weight is sizable;
for instance, it is
%about $1.2$
on the order of $1$
$\mu_B$/Fe in CaFe$_2$As$_2$ \cite{Zhao}.
These properties cannot be accounted for by the electrons close
to the Fermi energy alone;
%, especially given that
in particular,
since the Fermi surfaces
comprise small electron and hole pockets, the spin spectral weight
coming from the electronic states near the Fermi surfaces will be much smaller
than that observed experimentally.
Instead, the observed spin excitation spectrum is
%The observed properties}
%They instead
%are
more naturally associated with the electronic
states far away from the Fermi energy, as would be the case
in a metal
close to a Mott localization.

The incipient Mott picture arises when $U/t$ is not too far away from
the critical value for a Mott transition. (Here $t$
refers to the
%and $U$ are respectively the
characteristic bandwidth of the Fe 3d electrons,
and $U$ a combination of their Coulomb repulsions and Hund's couplings.)
%and Coulomb repulsion of the Fe 3d electrons.)
In order to further substantiate this picture, it is important to
tune the system into a Mott insulating state.
Recently, this has been demonstrated \cite{Zhu10}
in the iron oxychalcogenides, which contain an Fe square lattice
that is expanded compared to the iron pnictides and iron
chalcogenides. The lattice expansion gives rise to a narrowing of the
3d-bands and a concomitant enhancement of $U/t$, which pushes
the system through the Mott transition
and into the Mott-insulating regime.

Based on
the
above considerations, it is very important to show theoretically
that a transition from a metal to a Mott insulator generally exists in multi-orbital models
appropriate for
the
parent compounds of the iron pnictides.
This is especially so given that the number of 3d-electrons
per Fe is even in these systems.
The Mott transition, which has long been a subject
of fundamental interest~\cite{ImadaRMP98},
is studied in both
one-orbital Hubbard model with one electron
per site~\cite{DagottoRMP94,GKKRRMP96}, and multi-orbital systems at commensurate fillings~\cite{ImadaRMP98}.
In the one-orbital model and paramagnetic phases, the Mott transition
% is
can be understood
within the Brinkman-Rice
picture~\cite{BrinkmanRice}, with an interaction-induced suppression
of the coherent one-electron spectral weight near the Fermi energy,
and the concomitant development of incoherent spectral weight away
from the Fermi energy~\cite{GKKRRMP96}. Focusing on the paramagnetic
phases is advantageous for considering the Mott transition, since an
odd number of electrons per unit cell means that any insulating state
must be the result of interactions. The picture applies in the
part of the phase diagram above the temperature of ordering transitions
(typically into antiferromagnetic phases); this part of the
phase diagram widens with the increase of magnetic frustration
~\cite{Ohashietal08}.

In our case, since the parent iron pnictides contain
an even number of (six) 3d-electrons per Fe, an insulating state
could in principle simply be a band insulator. The issue is particularly
pertinent given that the Fermi surfaces of the iron pnictides and related
systems are small pockets.
Indeed, the non-interacting band structure consists of non-degenerate bands
(%in this paper,
 two bands
 are called
 degenerate
only when their energies are
% degenerate
the same
at every point in the momentum space) and is close to that of
a semiconductor
with a small overlap in energy between the bottom of the conduction bands
and the top of the valence bands;
each band is far away from half-filling.
It is {\it a priori} possible that, if an increasing $U$ suppresses
a metallic state, the system goes into a correlated band insulator first,
thereby invalidating the picture of a bad metal on the verge
of a Mott transition.
In other words, it is a non-trivial question
in the case of the iron pnictides as to whether
the metallic state can be in
proximity to a Mott transition and, if so,
whether the Mott insulator resembles that
for the canonical
one-orbital Hubbard model
with
%one-electron
one electron per unit cell.
Moreover, given that the bandwidths are very different from band to band, another non-trivial question is
%if a Mott transition exists, whether it
whether a Mott transition, when it exists,
is a one-step transition or it can be an orbital selective Mott transition (OSMT).\cite{Anisimovetal02}

To address
%above
these
issues, we study how a metal-insulator transition (MIT) in the
paramagnetic phase may happen in
% a multi-orbital model with
multi-orbital models of iron pnictides containing
an even
number of electrons filled.
Our work builds on earlier studies of multi-orbital models
for transition
metal oxides and related systems.
Physics in these multi-orbital models is very rich. For instance,
%a question that has been intensively discussed is whether an OSMT may occur.~\cite{Anisimovetal02}
%It has been shown that
many factors,
such as inequivalent bandwidths, crystal field splitting, and Hund's
coupling, may affect the nature of the transition
~\cite{Liebsch03,Kogaetal04,Knechtetal05,AritaHeld05,Werneretal09,Leeetal0910}.
We note that there
%There
is a tendency
in the literature
to work in the band representation and
ignore the orbital characters.\cite{Onoetal03,FlorensGeorges04}
In our study, we show that the orbital characters serve as another important factor
for the Mott transition.

We first study the MIT
in a two-orbital Hubbard model with two electrons per site,
with a kinetic-energy part given by the minimal band dispersion
for the pnictides~\cite{Raghu}. We analyze the model within
a slave-spin (SS) formulation~\cite{deMedici05,deMedici10},
which has the advantage of readily capturing
both the coherent and incoherent part of the electronic spectrum
and, in addition, the effect of Hund's coupling.
We show that the
orbital degeneracy guarantees the existence of a one-step metal-to-Mott-insulator transition
in this model. Neither a band insulator nor an OSMT may take place. The critical coupling $U_c$, which is larger in
multi-orbital systems than in the single-orbital
case \cite{Lu,Rozenberg}, is greatly reduced
by a
%finite
nonzero
Hund's coupling; this is consistent with earlier
studies in other multi-orbital contexts.\cite{HundRule}
We find that the nature of the Mott
insulator depends on the ratio of the inter-orbital and intra-orbital
Coulomb repulsions. It can be either a high-spin Mott state which is driven
by the intra-orbital coupling, or a low-spin orbital-Mott state driven
by the inter-orbital coupling.

We then consider a more realistic four-orbital model including
$xz$, $yz$, $xy$, and $x^2-y^2$
orbitals. In this model, the nature of the MIT depends on the strength of the Hund's coupling. A strong Hund's coupling stabilizes
a high-spin Mott state on the insulator side. But when the Hund's
coupling is either zero or weak compared to the crystal field splitting, we find a novel orbital selective MIT. The $xz$
and $yz$ orbitals
experience
a transition to a Mott insulator;
at the same critical coupling, a transition to either a band insulator (at zero Hund's coupling) or an orbitally polarized insulator (at
%finite
nonzero
albeit weak Hund's coupling)
takes place in the $xy$ and $x^2-y^2$ orbitals. In this case, the insulating state always has an intermediate spin value, even at zero Hund's coupling. We
%show this is
establish this to be
a direct consequence of the double degeneracy of the $xz$ and $yz$ orbitals.

%Our work builds on earlier studies of multi-orbital models
%for transition
%metal oxides and related systems.
%Physics in these multi-orbital models is very rich. For instance,
%a question that has been intensively discussed is whether an orbital
%selective Mott transition (OSMT) may occur.~\cite{Anisimovetal02}
%It has been shown that many factors,
%such as inequivalent bandwidths, crystal field splitting, and Hund's
%coupling, may affect the nature of the transition
%~\cite{Liebsch03,Kogaetal04,Knechtetal05,AritaHeld05,Werneretal09,Leeetal0910}.
%There is a tendency to work in the band representation and
%ignore the orbital characters, which could be another important factor
%for the Mott transition.

The
remainder of the
paper is organized as follows. In Sec. II we summarize the
slave spin formulation and introduce the two-orbital model.
Secs. III is devoted to the MIT in the two-orbital model
at half-filling. In particular, we propose the concept of
low-spin orbital-Mott state.
%, which may be connected adiabatically to a band insulator.
The investigation of MIT
in the four-orbital model is presented in Sec. IV, where a rich phase diagram is given.
Sec. V contains some
concluding remarks.

\section{Multi-orbital Hubbard Model and Slave-spin Representation}
The Hamiltonian for the multi-orbital Hubbard model
is
$H=H_0+H_{int}$. Here, $H_0$ is a non-interacting tight-binding
Hamiltonian with the general form
\begin{equation}\label{Ham_0}
H_0 = \sum_{\alpha\beta\mathbf{\nu}} t_{\alpha\beta}^{\mathbf{\nu}}
\sum_{\mathbf{i},\sigma} d^\dagger_{\mathbf{i}\alpha\sigma}
d_{\mathbf{i}+\mathbf{\nu}\beta\sigma} %+ \rm{H.c.}
+ \sum_{\mathbf{i},\alpha\sigma} (\epsilon_\alpha-\mu) d^\dagger_{\mathbf{i}
\alpha\sigma}d_{\mathbf{i}\alpha\sigma}
\end{equation}
in real space, where $d^\dagger_{\mathbf{i}\alpha\sigma}$ creates an electron on site $\mathbf{i}$, in orbital $\alpha$, and with spin $\sigma$.
$\epsilon_\alpha$ is the on-site potential of orbital $\alpha$ that
incorporates the crystal field splitting. $\mu$ is the chemical potential
determined by the electron filling. The interaction reads
\begin{eqnarray}\label{Ham_int}
H_{int} &=& U\sum_{\mathbf{i}\alpha} n_{\mathbf{i}\alpha\uparrow} n_{\mathbf{i}\alpha\downarrow} +(U'-J/2)\sum_{\mathbf{i},\alpha<\beta} n_{\mathbf{i}\alpha} n_{\mathbf{i}\beta} \nonumber\\
&-& J\sum_{\mathbf{i},\alpha<\beta} \left[ 2\mathbf{S}_{\mathbf{i}\alpha}\cdot \mathbf{S}_{\mathbf{i}\beta} - (d^\dagger_{\mathbf{i}\alpha\uparrow}d^\dagger_{\mathbf{i}\alpha\downarrow} d_{\mathbf{i}\beta\downarrow}d_{\mathbf{i}\beta\uparrow} +\rm{h.c.})\right] \nonumber \\
\end{eqnarray}
in which $U$($U'$) denotes the intra-(inter-)orbital Coulomb repulsion
and $J$ the Hund's coupling. These three parameters satisfy $U'=U-2J$ based on the consideration of rotational symmetry. It is strictly satisfied for an isolated atom in free space where all the d-orbitals are degenerate and the Coulomb potential has the full rotational symmetry, and is assumed to be also valid in solids.\cite{Castellani}
We will
% accept
adopt
this widely
% quoted
used
relation unless otherwise specified (see discussion in Sec.IIIB).
The spin operators are $\mathbf{S}_{\mathbf{i}\alpha} = \frac{1}{2}
\sum_{\sigma\sigma'} d^\dagger_{\mathbf{i}m\sigma} \vec{\tau}_{\sigma\sigma'}
d_{\mathbf{i}m\sigma'}$, where $\vec{\tau}=(\tau^x,\tau^y,\tau^z)$ are
the Pauli matrices.

It is now generally accepted that the degenerate $d_{xz}$ and $d_{yz}$
orbitals contribute most to the low-energy physics of the parent
iron pnictides. Hence in this paper, we will first consider a two-orbital
model introduced in Ref.~\onlinecite{Raghu}. The simplicity of the model makes it
easier to bring out some essential insights, which will also be
instructive for the understanding of more realistic models with
a larger number of orbitals.
Defining $\psi^\dagger_{\mathbf{k}\sigma} = (d_{\mathbf{k}x\sigma},
d_{\mathbf{k}y\sigma})$, we have $H_0 = \sum_{\mathbf{k}\sigma}
\psi^\dagger_{\mathbf{k}\sigma} [(\varepsilon_+-\mu)\mathbf{1}
+\varepsilon_-\tau^z+\varepsilon_{xy}\tau^x] \psi_{\mathbf{k}\sigma}$,
where $\varepsilon_+$, $\varepsilon_-$, and $\varepsilon_{xy}$ are the
intra- and inter-orbital hopping matrices in the momentum space.
%\ry{Compare to Eq.~\ref{Ham_0} we see that the orbital degeneracy requires $\epsilon_\alpha=0$.}
In the notation of Eq.~\ref{Ham_0}, the orbital degeneracy
requires an $\alpha$-independent
$\epsilon_\alpha$ which can be set to zero.
%\qs{QS NOTE?: RONG, I FIND THE PREVIOUS SENTENCE A BIT AMBIGUOUS; SEE WHETHER MY CHANGE IS OK.
We also notice that the tight-binding Hamiltonian in this model
is symmetric under the orbital interchange $xz\leftrightarrow yz$.
The parent compound has a half-filling, i.e., two electrons
per site.

We study the MIT in this two-orbital model
using the
SS formulation.\cite{deMedici05,deMedici10}
This formulation involves a much smaller number of slave fields
compared to the atomic-configuration-based slave-boson
representation of Ref.~\onlinecite{KotliarRuckenstein},
and more readily treats the full Hund's coupling compared to the
slave-rotor representation of Refs.~\onlinecite{FlorensGeorges04}
and \onlinecite{FlorensGeorges02}.
%In this approach,
Here,
a slave quantum $S=1/2$
spin is introduced on each site
for each orbital and spin degree of freedom:
$d_{\mathbf{i}\alpha\sigma}\rightarrow 2S^x_{\mathbf{i}\alpha\sigma}
f_{\mathbf{i}\alpha\sigma}$. The Hilbert space spanned by the SS and
auxiliary fermions are limited to the physical part by imposing
the constraint $S^z_{\mathbf{i}\alpha\sigma}+1/2 = n_{\mathbf{i}
\alpha\sigma}$ on each site.
The SS formulation handles the electron interactions by rewriting
$H_{int}$ in terms of SS operators. The
density-density
interactions
in Eq.~\ref{Ham_int}
(including the Ising-type Hund's coupling) are easily handled by the
z-component of the SS operator. The spin-flip part of the Hund's
coupling and the pair-hopping term are approximately treated
by substituting the fermion operators by SS operators that have
the same effect on the SS quantum numbers of the Hilbert space,
%: namely,
{\it viz.}
$-J\sum_{\mathbf{i}}
\left[S^+_{\mathbf{i}1\uparrow} S^-_{\mathbf{i}1\downarrow}
S^+_{\mathbf{i}2\downarrow} S^-_{\mathbf{i}2\uparrow} -
S^+_{\mathbf{i}1\uparrow} S^+_{\mathbf{i}1\downarrow}
S^-_{\mathbf{i}2\downarrow} S^-_{\mathbf{i}2\uparrow} +
\rm{H.c.}\right]$.
This
approximation
should capture the qualitative physics
because the SU(2) spin-rotational symmetry of the Hund's
coupling is still preserved in the Hilbert space spanned
by the slave spins.\cite{SU2}

The SS formulation is treated at the mean-field (MF) level
by fully decoupling the SS and auxiliary fermion operators
via a
saddle-point approximation.
This leads to two decoupled MF Hamiltonians for the SS
and the auxiliary fermions:
\begin{eqnarray}\label{e.hspin1}
H^S &=& 4\sum_{\alpha\beta\sigma}\sum_{\mathbf{i}\nu} S^x_{\mathbf{i},\alpha\sigma} S^x_{\mathbf{i}+\nu,\beta\sigma} \langle t^\nu_{\alpha\beta}f^\dagger_{\mathbf{i},\alpha\sigma}f_{\mathbf{i}+\nu,\beta\sigma}\rangle \nonumber\\
&+& \sum_{\mathbf{i},\alpha\sigma} h_{\alpha\sigma} (S^z_{\mathbf{i},\alpha\sigma}+1/2) + H^S_{int},
\end{eqnarray}
\begin{eqnarray}\label{e.hfermion1}
H^f &=& 4\sum_{\alpha\beta\sigma}\sum_{\mathbf{i}\nu} \langle S^x_{\mathbf{i},\alpha\sigma} S^x_{\mathbf{i}+\nu,\beta\sigma}\rangle t^\nu_{\alpha\beta}f^\dagger_{\mathbf{i},\alpha\sigma}f_{\mathbf{i}+\nu,\beta\sigma}\nonumber\\
&+& \sum_{\mathbf{i}\alpha\sigma} (\epsilon_\alpha-h_{\alpha\sigma}-\mu) f^\dagger_{\mathbf{i},\alpha\sigma} f_{\mathbf{i},\alpha\sigma};
\end{eqnarray}
where
\begin{eqnarray}\label{e.hsint}
H^S_{int} &=& \sum_{\mathbf{i}} \left\{\frac{U'}{2}\left(\sum_{\alpha\sigma} S^z_{\mathbf{i}\alpha\sigma}\right)^2 + \frac{U-U'}{2} \sum_{\alpha}\left(\sum_{\sigma} S^z_{\mathbf{i}\alpha\sigma}\right)^2\right. \nonumber\\ &-& \frac{J}{2} \sum_{\sigma}\left(\sum_{\alpha} S^z_{\mathbf{i}\alpha\sigma}\right)^2 -J
\left[S^+_{\mathbf{i}1\uparrow} S^-_{\mathbf{i}1\downarrow}
S^+_{\mathbf{i}2\downarrow} S^-_{\mathbf{i}2\uparrow} \right. \nonumber\\
&-&\left.\left.
S^+_{\mathbf{i}1\uparrow} S^+_{\mathbf{i}1\downarrow}
S^-_{\mathbf{i}2\downarrow} S^-_{\mathbf{i}2\uparrow} +
\rm{H.c.}\right]\right\},
\end{eqnarray}
and $h_{\alpha\sigma}$ is a Lagrangian multiplier taking account for the constraint. To solve these two Hamiltonians, we further apply the mean-field decomposition to the term $S^x_{\mathbf{i},\alpha\sigma} S^x_{\mathbf{i}+\nu,\beta\sigma}$ in $H^S$, i.e., $S^x_{\mathbf{i},\alpha\sigma} S^x_{\mathbf{i}+\nu,\beta\sigma} \approx \langle S^x_{\mathbf{i},\alpha\sigma}\rangle S^x_{\mathbf{i}+\nu,\beta\sigma}+ S^x_{\mathbf{i},\alpha\sigma} \langle S^x_{\mathbf{i}+\nu,\beta\sigma}\rangle - \langle S^x_{\mathbf{i},\alpha\sigma}  S^x_{\mathbf{i}+\nu,\beta\sigma}\rangle$, and assume $\langle S^x_{\mathbf{i},\alpha\sigma}\rangle$
%is
to be
site independent. The Hamiltonian for the SS operators is then reduced to (up to a constant)
\begin{eqnarray}\label{e.hspin2}
H^S_{MF} &=& \sum_{\mathbf{i},\alpha\sigma} \left[ K_{\alpha\sigma} S^x_{\mathbf{i},\alpha\sigma}
+ h_{\alpha\sigma} (S^z_{\mathbf{i},\alpha\sigma}+1/2) \right] + H^S_{int},
\nonumber\\
\end{eqnarray}
where $K_{\alpha\sigma}=8\sum_{\beta} \langle S^x_{\beta\sigma} \rangle \sum_{\mathbf{i}\nu}\langle t^\nu_{\alpha\beta} f^\dagger_{\mathbf{i},\alpha\sigma}f_{\mathbf{i}+\nu,\beta\sigma} \rangle$.
The quasiparticles near
the Fermi level are described by the auxiliary fermion Hamiltonian $H^f$. Introducing the quasiparticle spectral weight
$Z_\alpha=4\langle S^x_{\alpha\sigma}\rangle^2$,
% at the MF level,
$H^f$ is written as
\begin{eqnarray}\label{e.hfermion2}
H^f_{MF} &=& \sum_{\alpha\beta\sigma}\sum_{\mathbf{i}\nu} \sqrt{Z_\alpha Z_\beta} t^\nu_{\alpha\beta}f^\dagger_{\mathbf{i},\alpha\sigma}f_{\mathbf{i}+\nu,\beta\sigma}\nonumber\\
&+& \sum_{\mathbf{i}\alpha\sigma} (\epsilon_\alpha-h_{\alpha\sigma}-\mu) f^\dagger_{\mathbf{i},\alpha\sigma} f_{\mathbf{i},\alpha\sigma};
\end{eqnarray}
which has a similar form
%to
as
 $H_0$, with the hopping $t_{\alpha\beta}^{\nu}$
renormalized to $\sqrt{Z_\alpha Z_\beta} t_{\alpha\beta}^{\nu}$.
In practice, Eq.~\ref{e.hspin2} and Eq.~\ref{e.hfermion2} are
self-consistently solved by iteratively determining the
parameters $h_{\alpha\sigma}$ and $K_{\alpha\sigma}$ (hence $Z_\alpha$).
The metallic behavior corresponds to the Bose condensation
of the slave spins, which is marked by a non-zero $Z_\alpha$.
The Mott insulating behavior of orbital $\alpha$ is then
identified by $Z_\alpha=0$.

\begin{figure}[th]
\includegraphics[width=1.0\linewidth]{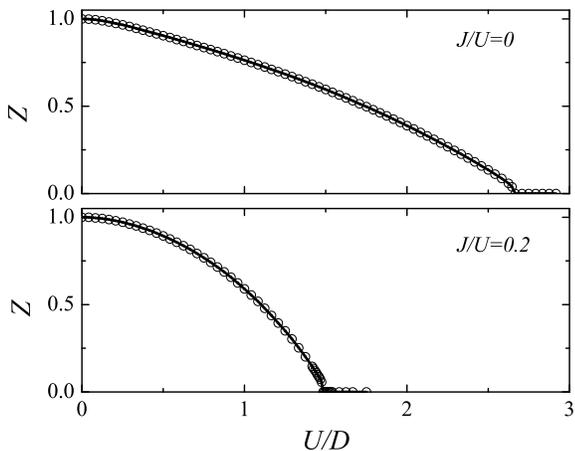}
%\vskip -1.0cm
\caption{(Color online) Evolution of the spectral weight $Z$ in the two-orbital model with $J=0$, $U'=U$ in (a) and $J/U=0.2$, $U'/U=0.6$ in (b). $D=12$, is the full bandwidth of the non-interacting band structure.
}
\label{f.fig1}
\end{figure}

\begin{figure}[th]
\includegraphics[width=1.0\linewidth]{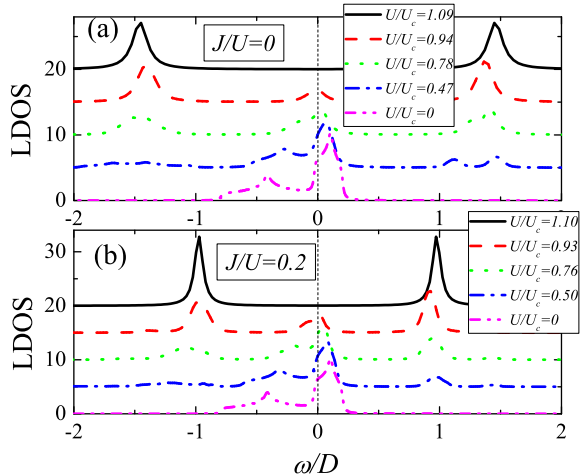}
%\vskip -1.0cm
\caption{(Color online) DOS in the two-orbital model at various $U$ values with the same model parameters as in Fig.~\ref{f.fig1}.
}
\label{f.fig2}
\end{figure}

\section{Results
% in
for
the two-orbital model}
\subsection{Mott transition in the two-orbital model}

In the two-orbital model for iron pnictides, the non-interacting tight-binding Hamiltonian contains an inter-orbital
hopping $\varepsilon_{xy}$, and cannot be diagonalized by a uniform
($\mathbf{k}$-independent) orbital rotation. This leads to two non-degenerate bands which has asymmetric local density of states (LDOS) with respect to the chemical potential. The Fermi surface consists small electron and hole
pockets, indicating that both bands are far away from half-filling.
It is then important to ask
whether,
when the interactions are turned on,
%whether
the system
%may go through
undergoes
a
%MIT
transition
to a Mott insulator, or the band degeneracy can be fully lifted so that a correlated band insulator may be stabilized. If a Mott transition
%can
does
take place, it is also interesting to ask whether the transition
to the Mott insulator is
%via
through
a single MIT or
via an OSMT.

Most of previous studies on the MIT in multi-orbital
systems work in the band presentation,
%namely, assume
assuming
a diagonalized
band structure with zero inter-band hopping. This oversimplified treatment
neglects the orbital character of the model.
Within this representation, the only way to generate two bands with asymmetric LDOS is to introduce a finite crystal field
splitting which
%already
breaks the orbital degeneracy in the model.
In this study we choose to work in the orbital representation, which takes
% account for
into account
the full orbital characters of the model. In this orbital representation, the symmetry of $H_0$ and $H_{int}$ under interchanging the $xz$ and $yz$ orbitals guarantees that both orbitals contribute equally to the band structure.

The above considerations lead to two important consequences. First,
$n_{xz,\sigma}=n_{yz,\sigma}=1/2$ for any value of $U$, i.e.,
each orbital is exactly at half-filling. So a Mott transition is
possible at finite $U$. Second, $Z_{xz} = Z_{yz} = Z$, and an OSMT
cannot take place
%although the bandwidths
even though the widths of the two bands
%\qs{(QS NOTE?: RONG CAN YOU CHECK THAT MY REWORDING HERE IS OKAY?)}
are very different. At finite $U$, the hopping parameter $t_{lm}^{\nu}$ is
uniformly renormalized to $\sqrt{Z_lZ_m} t_{lm}^{\nu} = Zt_{lm}^{\nu}$ since
the crystal field splitting is zero in this model. Hence the dispersion
$\epsilon_{\mathbf{k}}$ (the Fourier transformation of $t_{lm}^{\nu}$) is
also normalized to $Z\epsilon_{\mathbf{k}}$. We thus see that the topology
of the band structure is unchanged
%in the presence of
by the interactions. At $T=0$, the chemical potential is determined by $\int \Theta(\mu-Z\epsilon_{\mathbf{k}})=2N_{s}$, where $\Theta(x)$ is the Heaviside step function, and $N_s$ the number of sites. Taking $Z=1$ and $\mu=\mu_0$ at $U=0$, we see that $\int \Theta(\mu-Z\epsilon_{\mathbf{k}})=\int \Theta(Z(\mu_0-\epsilon_{
 \mathbf{k}}))$,
%implies
implying
that $\mu=Z\mu_0$. The Fermi surface is determined by
$Z(\mu_0-\epsilon_{\mathbf{k}})=0$. Therefore, the Fermi surface of the
interacting system is \textit{identical} to the one in the non-interacting
system.
%It also indicates that
Furthermore,
the filling factor of each quasi-particle band does not depend on the
interaction and must be identical to the value in the non-interacting case.
Therefore, in the presence of
%finite
nonzero
interaction, the quasi-particle bands are still partially occupied, and a band insulator never emerges. The system stays in the metallic state until the quasi-particle spectral weight $Z=0$, where both bands go through a transition to a Mott insulator. Our argument generally applies to any system with degenerate orbitals.

The above analysis is supported by the full solution to the
slave-spin mean-field (SSMF) equations.
Fig.~\ref{f.fig1} shows the evolution of spectral weight $Z$
%as
with
increasing
$U$. For both $J=0$ and
%finite
nonzero
$J$, $Z$ drops to zero at finite $U$,
indicating a Mott transition. At $J=0$, this takes place at $U_c/D=2.66$,
where $D=12$ is the full bandwidth of the two-orbital model (in the unit
of $t_1$, the nearest-neighbor intra-$xz$-orbital hopping along the $x$
direction).
The critical
coupling is reduced to $U_c/D=1.49$ at $J/U=0.2$.
As a complementary method, we have also applied
%We also apply
the slave-rotor formulation~\cite{FlorensGeorges04} to the two-orbital model with
J=0 and the same
values for the other
model parameters. The results (not shown) are qualitatively the same to the SSMF results at $J=0$: a Mott transition takes place at $U_c/D\approx2.0$.
%This confirms that a Mott transition exists in the two-orbital model.

The Mott transition is best seen in the variation of the spectral function
%as
with
increasing $U$. We calculate the spectral function by convoluting the SS and auxiliary fermion Green's functions: $G^{rd}_{\alpha\beta}(\mathbf{k},\omega) = i\int d\omega'\{G^{>S}_{\alpha\beta}(\omega') G^{rf}_{\alpha\beta}(\mathbf{k},\omega-\omega') + G^{rS}_{\alpha\beta}(\omega') G^{<f}_{\alpha\beta}(\mathbf{k},\omega-\omega')\}$, where $G^r$ is the retarded Green's function, $G^{>S}_{\alpha\beta}(t)\equiv -i\langle S^x_\alpha(t)S^x_\beta(0)\rangle$, and $G^{<f}_{\alpha\beta}(t)\equiv i\langle f^\dagger_\beta(0)f_\alpha(t)\rangle$ and $\alpha,\beta$ denote orbital indices. Note that in
the
above expression the SS Green's function is independent of $\mathbf{k}$. This is a consequence of the
MF approximation. Expressing the SS Green's function using
the
Lehmann representation and taking
into
account that the Hamiltonian of $f$-fermions describes free fermions, the spectral function is written as
\begin{eqnarray}\label{Akw}
A(\mathbf{k},\omega) &=& \frac{2\pi}{\mathcal{Z}} \sum_{\alpha\lambda\sigma}\sum_{n,m} \left|\langle n|S^x_{\alpha\sigma}|m\rangle\right|^2 |\mathbf{\Lambda}^{\alpha\lambda}_\mathbf{k}|^2 \delta(\omega-E_{nm}-\epsilon_{\lambda\mathbf{k}}) \nonumber\\
&\times& \left\{e^{-\beta E_m}(1-n^f_{\lambda\mathbf{k}})+e^{-\beta E_n}n^f_{\lambda\mathbf{k}}\right\},
\end{eqnarray}
where $E_n$ and $|n\rangle$ are eigenenergy and eigenvector of SS Hamiltonian, $\mathcal{Z}=\sum_n e^{-\beta E_n}$,
%and
 $E_{nm}=E_n-E_m$,
and $n^f_\lambda=1/(e^{\beta\epsilon_\lambda}+1)$.
The matrix $\mathbf{\Lambda}_\mathbf{k}$ diagonalize the
$f$-fermion Hamiltonian $H^f_{MF}$ with eigenenergy
$\epsilon_{\lambda\mathbf{k}}$.
 %$n^f_\lambda=1/(e^{\beta\epsilon_\lambda}+1)$.
Taking $m=n=0$ ($|0\rangle$ denotes the ground state of $H^S_{MF}$) in Eq.~\ref{Akw}, one sees that the coherent part of the spectral function is normalized by a factor of $Z$ since $Z=|\langle 0|S^x_{\alpha\sigma}|0\rangle|^2$.
As mentioned above, an advantage of the SSMF is that the incoherent part is also accessible. At low temperatures this comes from terms with $n=0,m\neq0$. The LDOS is calculated from Eq.~\ref{Akw}. As shown in Fig.~\ref{f.fig2}, when $U$ is increased from zero, one sees clearly that the coherent part is renormalized by $Z$. There is a significant spectral weight transfer to the incoherent part in the metallic phase.
%Hence with SSMF it is able to explain
Our results provide a natural explanation of
both the renormalization of the coherent bands and the appearance of the incoherent spectral weights~\cite{Qazilbash,Dingetal08} within a unified
framework. At $U>U_c$, the coherent peak vanishes,
%indicating the transition to a
signaling the Mott insulator state; the incoherent parts, at the same time,
develop into the lower and upper Hubbard bands.

\begin{figure}[h]
%\vskip +0.5cm
%\includegraphics[
%bbllx=80pt,bblly=30pt,bburx=730pt,bbury=550pt
%clip,width=40mm]{PhDJfa.eps}
\includegraphics[
bbllx=30pt,bblly=30pt,bburx=700pt,bbury=560pt
clip,width=80mm]{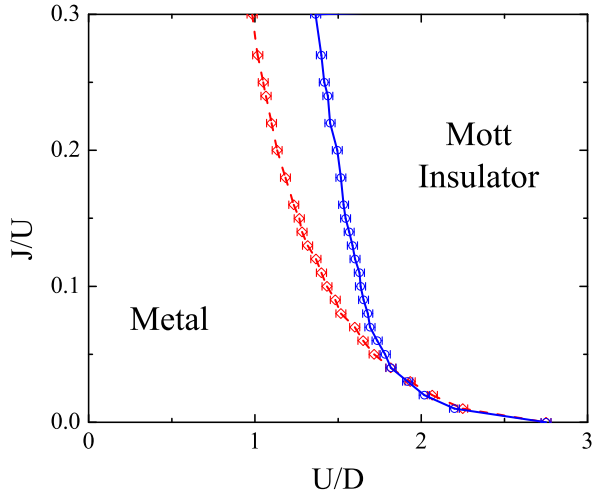}
\caption{(Color online) Phase diagram in the $J-U$ plane of the two-orbital model at half-filling.
Here we have taken $U'=U-2J$. The
% the
blue circles (red diamonds) give the phase boundary of
the metal-to-Mott-insulator transition
% if the
for the case of
 full (Ising-type) Hund's coupling.
% is considered.
}
\label{f.fig3}
\end{figure}

As discussed in previous studies,\cite{HundRule} a finite Hund's coupling
may strongly affect the Mott transition. We show this effect in the
two-orbital model by presenting the $J$-$U$ phase diagram in Fig.~\ref{f.fig3}.
%As
It is seen in Fig.~\ref{f.fig3} that $U_c$ is rapidly reduced with
increasing $J/U$;
%ratio from zero
%, $U_c$ is rapidly reduced,
% and this has also been seen
this is also illustrated
in Fig.~\ref{f.fig1}.
The reduction of $U_c$ can be understood by solving the mean-field Hamiltonian
in Eq.~\ref{e.hspin2} at an infinitesimal $K$ ($K\equiv K_{\alpha\sigma}$). We first diagonalize
$H_{int}$ and the term including $S^z_{\mathbf{i},\alpha\sigma}$ in Eq.~\ref{e.hspin2}, and label the eigenstates as $|n\nu\rangle$, where $n$ is the electron occupation number,
and $\nu$ denotes the degenerate multiplets.  We then treat the term
$KH'=K\sum_{\alpha\sigma} S^x_{\alpha\sigma}$ perturbatively.
To the first-order in $K$, $U_c$ can be obtained by solving $1/\bar{\epsilon}=2\cal{E}$.
Here $\bar{\epsilon}=1/N_{site}
\sum_{\alpha\beta} \epsilon^{\alpha\beta}_{\mathbf{k}}\langle f^\dagger_{\mathbf{k}\alpha\sigma}
f_{\mathbf{k}\alpha\sigma}\rangle$ is the average kinetic energy for the noninteracting system,
and $\cal{E}$ is the lowest eigenvalue diagonalizing the matrix $\mathbf{M}$,
where $\mathbf{M}_{\mu\nu}=\sum_{n\neq2,\lambda}
\langle 2\mu |H'| n\lambda \rangle \langle n\lambda |H'| 2\nu \rangle/(E_2-E_n)$ and $E_n$
is the eigenenergy of state $|n\nu\rangle$. $M_{\mu\nu}$ is non-zero only when $n=1$
or $n=3$. For either $n$ value, $E_2-E_n=-\Delta_2/2$ where $\Delta_2=U+J$ is the
Mott gap of the two-orbital model at half-filling. Hence $U_c$ is determined
by $\Delta_2\propto|\bar{\epsilon}|$, i.e., $U_c\propto|\bar{\epsilon}|/(1+J/U)$.
This clearly indicates that $U_c$ decreases with increasing $J/U$ ratio.
Similar behavior has also been discussed in a three-orbital model. \cite{Werneretal09}

Interestingly, we find that $U_c$ is reduced more significantly
for the Ising-type Hund's coupling (i.e., in the absence of the
spin-flip and pair-hopping terms in Eq.~\ref{e.hsint}) at large $J/U$ ratio, as shown in Fig.~\ref{f.fig3}.
This is quite consistent with the results in previous studies,\cite{HundRule} and can also be understood from the above perturbation theory. When $J$ is small, for either Ising-type or full Hund's coupling, all six
%$|2\mu\rangle$
configurations associated with two electrons occupying the two orbitals,
denoted by $|2\mu\rangle$ with $\mu$ ranging from 1 through 6,
are nearly degenerate, and are strongly mixed in the (perturbed) ground state. But when the full Hund's coupling $J$ is large, only the triplet configurations
in $|2\mu\rangle$,
shown in Fig.~\ref{f.fig4},
contribute most to the ground state. For
the
Ising-type Hund's coupling, the ground state only strongly mixes the doublet: $|\uparrow\rangle_{xz}|\uparrow\rangle_{yz}$ and
$|\downarrow\rangle_{xz}|\downarrow\rangle_{yz}$. More configurations mixed in the ground state correspond to more scattering processes between the nearly degenerate $|2\mu\rangle$ and $|2\nu\rangle$ states, which promote a larger kinetic-energy gain
in the metallic phase, thereby favoring the metal over
the Mott insulator. Following the perturbation theory, $U_c=12|\bar{\epsilon}|/(1+J/U)\approx2.7D/(1+J/U)$ for infinitesimal $J/U$;
here, a nearly degenerate perturbation is used involving all the six low-energy multiplets.
At
%large enough
sufficiently large
$J$, $U_c=8|\bar{\epsilon}|/(1+J/U)$ for full Hund's coupling and $U_c=4|\bar{\epsilon}|/(1+J/U)$ for Ising-type Hund's coupling;
here, the degenerate perturbation respectively
involves three and two lowest multiplets for the two cases.
These
expressions
are qualitatively ~\cite{Hund2} consistent with
the numerical results in Fig.~\ref{f.fig3}:
%, namely
%a
At $J/U<0.01$, $U_c\approx2.66D/(1+J/U)$ for both full and Ising-type Hund's couplings;
While for $J/U\gtrsim 0.03$ the largest reduction of $U_c$ is found in the Ising-type Hund's coupling.
%,in which case the kinetic energy gain is the least.~\cite{Hund2}
It is interesting to note that the effect of Hund's coupling on the value of $U_c$ is quite similar to the effect of having more orbitals in degenerate multi-orbital Hubbard models.\cite{Gunnarsson96} %Actually
Indeed
 %they share
 % the same physics behind
the underlying physics is related:
in both cases, $U_c$ is higher when the ground state mixes more
 %near degenerate
nearly-degenerate configurations;
as already mentioned,
 % because
involving more configurations helps stabilizing a metallic state
%  via lowering
by lowering the kinetic energy.

Another difference from the $J=0$ case is that the MIT becomes discontinuous
at
%finite
nonzero
$J$. It is especially significant for the Ising-type Hund's
interaction. For the full Hund's coupling, the discontinuity of $Z$ is only
significant at small $J$ values. For $J/U\gtrsim 0.2$, the discontinuity is
rather small (see Fig.~\ref{f.fig1}), and it is hard to distinguish
%it
the transition from a continuous one.
These results from our SSMF calculation are consistent with
the DMFT results on the effect of finite Hund's couplings
in multi-band Hubbard models with degenerate bands.~\cite{HundRule}
%\ryr{The discontinuous nature of the MIT also explains the discrepancy between the numerical %findings and results based on above perturbation theory at $J>0$.~\cite{Hund2}

\subsection{High-spin-Mott vs. low-spin orbital-Mott state}
\begin{figure}[h]
\includegraphics[
%bbllx=80pt,bblly=190pt,bburx=590pt,bbury=480pt
clip
,width=80mm]
%[width=1.0\linewidth]
{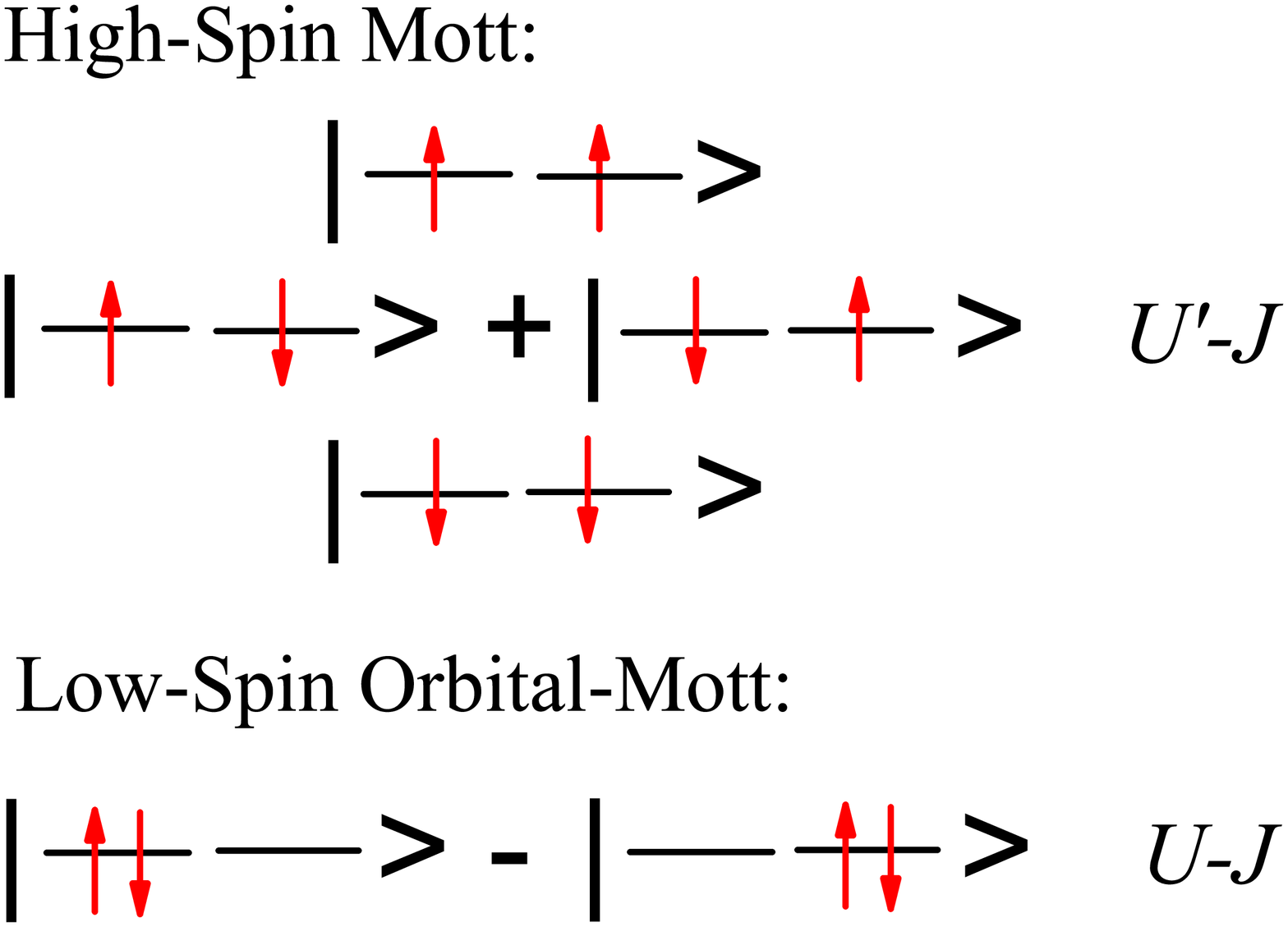}
\caption{(Color online) Illustration of the high-spin Mott and low-spin orbital-Mott states in the two-orbital model at the atomic limit. In this model, the high-spin Mott state has $S=1$ and the ground-state energy $U'-J$; the low-spin orbital-Mott state has $S=0$ and the ground-state energy $U-J$.
}
\label{f.fig4}
\end{figure}
As seen in Fig.~\ref{f.fig2},
%the picture of
the Mott transition at half-filling in the two-orbital model is
quite similar to the Brinkman-Rice picture of the one-orbital case.
It is then interesting to see whether the Mott insulating state is similar
to that of the one-orbital case.
For $J>0$, the Mott insulator of the two-orbital model is an $S=1$ inter-orbital triplet state. This state is characterized by a vanishing on-site double occupancy $\langle n_{\mathbf{i}\uparrow} n_{\mathbf{i}\downarrow} \rangle$ in each orbital, and is the two-orbital analogue of the Mott insulator in the one-orbital model. We denote the triplet state as the high-spin Mott state in the two-orbital model.

The
Mott insulating
state at $J=0$ in the two-orbital model is
%a bit
somewhat
different. It mixes the high-spin (triplet) states with low-spin (singlet) configurations,
which are degenerate when $J=0$. Therefore, the insulating state
%and
has a finite double occupancy (see Fig.~\ref{f.fig4}).
This implies that a spin-singlet state might be stabilized in some parameter regime. Note that the relation $U'=U-2J$ puts a strong constraint
%in
on the parameters, which may limit the ground state configurations to
a relatively smaller subset. To fully study all the possible ground state
configurations, in this section, we relax the above constraint
so that $U'$ becomes a free parameter independent of
$U$ and $J$.\cite{note1,Yanagi10} Indeed, by studying the Hamiltonian in
the atomic limit, one sees that for any $J>0$ the inter-orbital triplet
state is only stabilized at $U'<U$. When $U'>U$, the ground state is
an orbital anti-symmetric spin-singlet state
$1/\sqrt{2}(|\uparrow\downarrow;0\rangle - |0;\uparrow\downarrow\rangle)$,
as shown in Fig.~\ref{f.fig4}.\cite{note} Though this state shares some
characters as a band insulator, such as finite double occupancy and
spin singlet, it is still a Mott insulator because the orbital degeneracy is
 %still
preserved by the Hamiltonian. This can be immediately
 %see
seen by noticing
that each orbital is at half-filling and the spectral weight is zero in this state. Note that $U'>U$ is not enough to drive the system to a band insulator because the existence of the Mott insulator is guaranteed by the orbital degeneracy. To distinguish this Mott state at $U'>U$ from the high-spin (triplet) Mott state,
we will denote it
 %it is denoted
as low-spin orbital-Mott state.
 % in this paper.}

%In the two-orbital model, as shown in Fig.~\ref{f.fig5}, for $U'<U$ the double
%occupancy drops to zero at $U=U_c$. But for $U'\geqslant U$, the double o
%occupancy remains finite even in the Mott insulator. This can be understood in
%the Mott insulator at the $U\rightarrow\infty$ limit. In the atomic limit, f
%or any $J>0$, the ground state is an inter-orbital spin triplet if $U'<U$.
%We denote the corresponding Mott insulator at finite $U$ as the spin-Mott
%state, as illustrated in Fig.~\ref{f.fig4}. If $U'>U$, the two electrons
%occupy the same orbital forming a singlet to avoid the penalty from the
%inter-orbital coupling. When the orbital degeneracy is broken, this
%corresponds to a band insulator, i.e., with one orbital double occupied,
%the other empty. But for degenerate orbitals the ground state
%is $1/\sqrt{2}(|\uparrow\downarrow;0\rangle + |0;\uparrow\downarrow\rangle)$.
%The average electron density in each orbital is still one per site,
%and the quasiparticle spectral weight is zero.
%Hence it has all the signature of a Mott insulator. To distinguish this state
%from the spin-Mott state, it is denoted as orbital-Mott state in this paper.

\begin{figure}[th]
\includegraphics[width=1.0\linewidth]{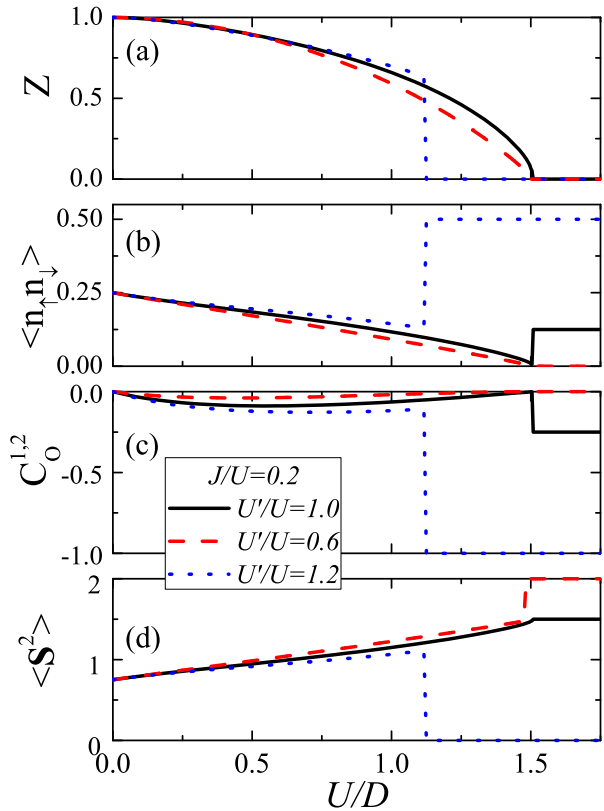}
\caption{(Color online) (a): The evolution of $Z$ at half-filling with $J/U=0.2$ and different $U'/U$ ratios showing the transition to different Mott states. (b)-(d): The evolution of $\langle n_\uparrow n_\downarrow\rangle$, $C_o^{1,2}$, and $\mathbf{S}^2$ (see text for the definitions of these quantities) for the same set of parameters.
}
\label{f.fig5}
\end{figure}

To see the difference between the high-spin Mott and low-spin orbital-Mott
insulators, we study the MIT in the two-orbital model at several different $U'/U$ ratios and show the results in Fig.~\ref{f.fig5}. We find the MIT is
discontinuous for general $U$, $U'$, and $J$ values except for $J=0$ and $U=U'$, where a continuous transition is observed. Moreover, the behaviors in the metallic state at different $U'/U$ ratios are quite similar. This is not surprising since the electron hopping mixes all configurations. Besides the double occupancy, the difference between the high-spin Mott and low-spin orbital-Mott states can also be
seen by both the average value of the total spin operator
$\mathbf{S}^2=(\sum_\alpha\mathbf{S}_\alpha)^2$, and the orbital correlation
function $C_o^{1,2} = \langle (n_1-1)(n_2-1)\rangle$. (We have defined
the orbital indices $1=xz$, $2=yz$.)
In the high-spin Mott state, $\langle\mathbf{S}^2\rangle=2$ and $C_o^{1,2}=0$.
By contrast, in the low-spin orbital-Mott state, $\langle\mathbf{S}^2\rangle=0$
and $C_o^{1,2}=-1$. All these are consistent with the numerical results
shown in Fig.~\ref{f.fig5}. At $U'=U$, the ground state is a mixture of the two Mott states, hence both $C_o^{1,2}$ and $\langle\mathbf{S}^2\rangle$ take intermediate values.

For a fixed $J/U$ ratio, we find the critical coupling $U_c$ for the
MIT is the largest for $U=U'$. It is slightly reduced for $U'<U$ but greatly decreased when $U'>U$. We show that this non-monotonic behavior of $U_c$ is related to the different nature of the Mott states. Note that the two Mott states at $U'<U$ and $U'>U$ have different Mott gaps. In the high-spin Mott state, $\Delta_2=U+J$, is independent of $U'$. On the other hand, the Mott gap in the low-spin orbital-Mott state is $\Delta_2=2U'-U+J$,
%it
which increases with $U'$.
%as U' increases.
Therefore, $U_c$ decreases drastically when $U'/U$ increases from 1, but
% $U_c$
is almost insensitive to $U'/U$ for $U'<U$.
%consistent with
These considerations allow us to understand
the numerical results given in Fig.~\ref{f.fig5}.\cite{note2}
The fact that $U_c$ at $U'<U$ is smaller than $U_c$ at $U'=U$ can be further understood by the different ground-state degeneracy in the two states. From Sec. IIIA we know that a higher ground-state degeneracy increases the effective kinetic energy gain, thereby
enhancing the stability of the metallic state. The ground state is three-fold degenerate when $U'<U$ but is four-fold degenerate when $U'=U$ (six-fold if further $J=0$). Thus, $U_c$ is the largest for $U=U'$, and is slightly reduced when $U'<U$.

The orbital-Mott state
exists only in systems with degenerate orbitals and $U'>U$.
When the orbital degeneracy
is broken by a crystal field splitting, it is unstable toward either a band insulator, or more generally as will be discussed in the next section, an orbitally polarized insulator.

\begin{figure}[h]
\includegraphics[
bbllx=70pt,bblly=190pt,bburx=720pt,bbury=480pt
clip
,width=80mm]
%[width=1.0\linewidth]
{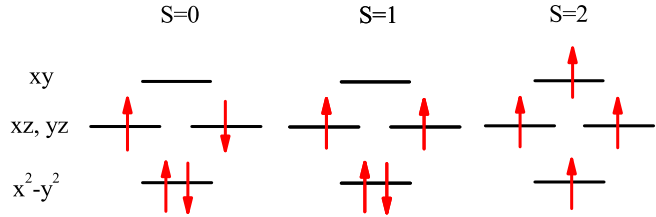}
\caption{(Color online) Three candidate ground states in the atomic limit in the four-orbital model. The $S=2$ state is an analogue of
the high-spin Mott state in the two-orbital model.
}
\label{f.Fig6}
\end{figure}

\section{Results for the Four-Orbital Model}
We have so far shown how the Mott
transition takes place in the minimal two-orbital model for parent iron
pnictides. It has been recognized that, to more realistically reproduce the
electron structure of the iron pnictides,
all the five Fe orbitals need to be included in the tight-binding dispersion
\cite{Graser,Kuroki}.
This raises the question as to
whether our main results for the two-orbital model are applicable to the
more realistic models with a larger number of orbitals.
In the five-orbital model of Ref.~\onlinecite{Graser}, the electron
filling is about $0.8$ per site per spin in the $3z^2-r^2$ orbital,
but very close to $0.5$ per site per spin in all other four orbitals.
The $3z^2-r^2$ orbital hardly contributes to the band structure
near the Fermi level. These suggest that one may study a model including only $xz$, $yz$, $xy$, and $x^2-y^2$ orbitals by assuming that the $3z^2-r^2$ orbital lies far below the Fermi level and is fully occupied.
%If taking
Taking
model parameters of Ref.~\onlinecite{Graser} but keeping
only those four orbitals
%, the resulting Fermi surfaces are
gives rise to Fermi surfaces that
almost identical
to the ones for the five-orbital cases. Hence the four-orbital model
represents a good approximation to the five-orbital one. For simplicity,
%in this paper
here
we study the MIT in this four-orbital model.

We argue that the main results for the two-orbital model still hold
in the four-orbital model.
Though the orbital degeneracy of other orbitals are lifted, the $xz$
and $yz$ orbitals are still degenerate. In the atomic limit, with four electrons occupied, the ground state may be either a high-spin $S=2$ state, or an intermediate-spin $S=1$ state, or a low-spin $S=0$ state, as shown in Fig.~\ref{f.Fig6}. In either case the $xz$ and $yz$ orbitals are half filled just as in the two-orbital model. Therefore, a Mott transition similar to the two-orbital case is expected.

\begin{figure}[h]
\includegraphics[
bbllx=20pt,bblly=60pt,bburx=710pt,bbury=590pt
clip
,width=80mm]
%[width=1.0\linewidth]
{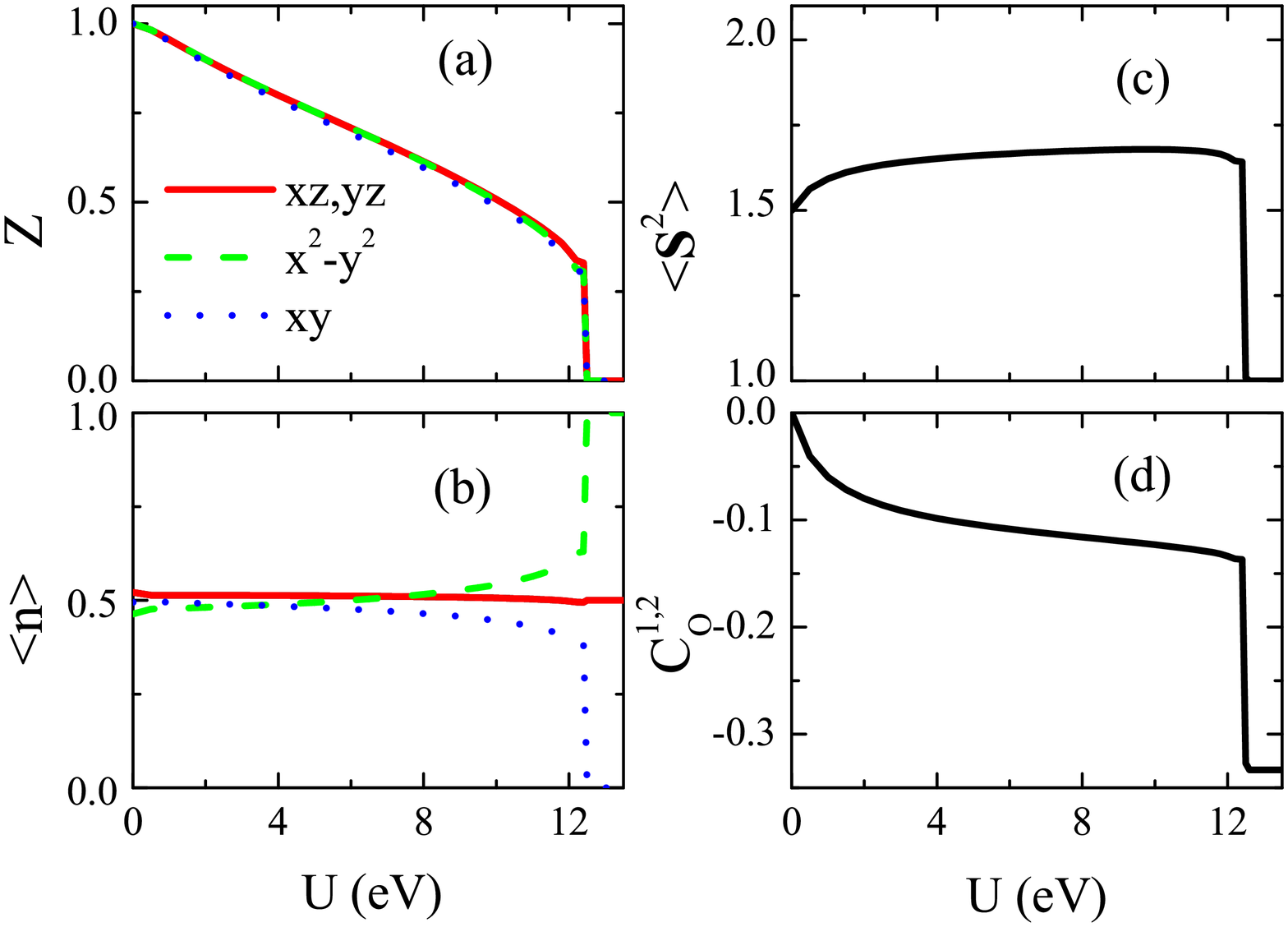}
\caption{(Color online) SSMF results for the four-orbital model at $J=0$, showing evolution of $Z$ (in (a)), average electron filling per site per spin (in (b)), average of total spin (in (c)), and inter-orbital correlation between $xz$ and $yz$ orbitals (in (d)).
}
\label{f.Fig7}
\end{figure}

\begin{figure}[h]
\includegraphics[
bbllx=20pt,bblly=60pt,bburx=710pt,bbury=590pt
clip
,width=80mm]
%[width=1.0\linewidth]
{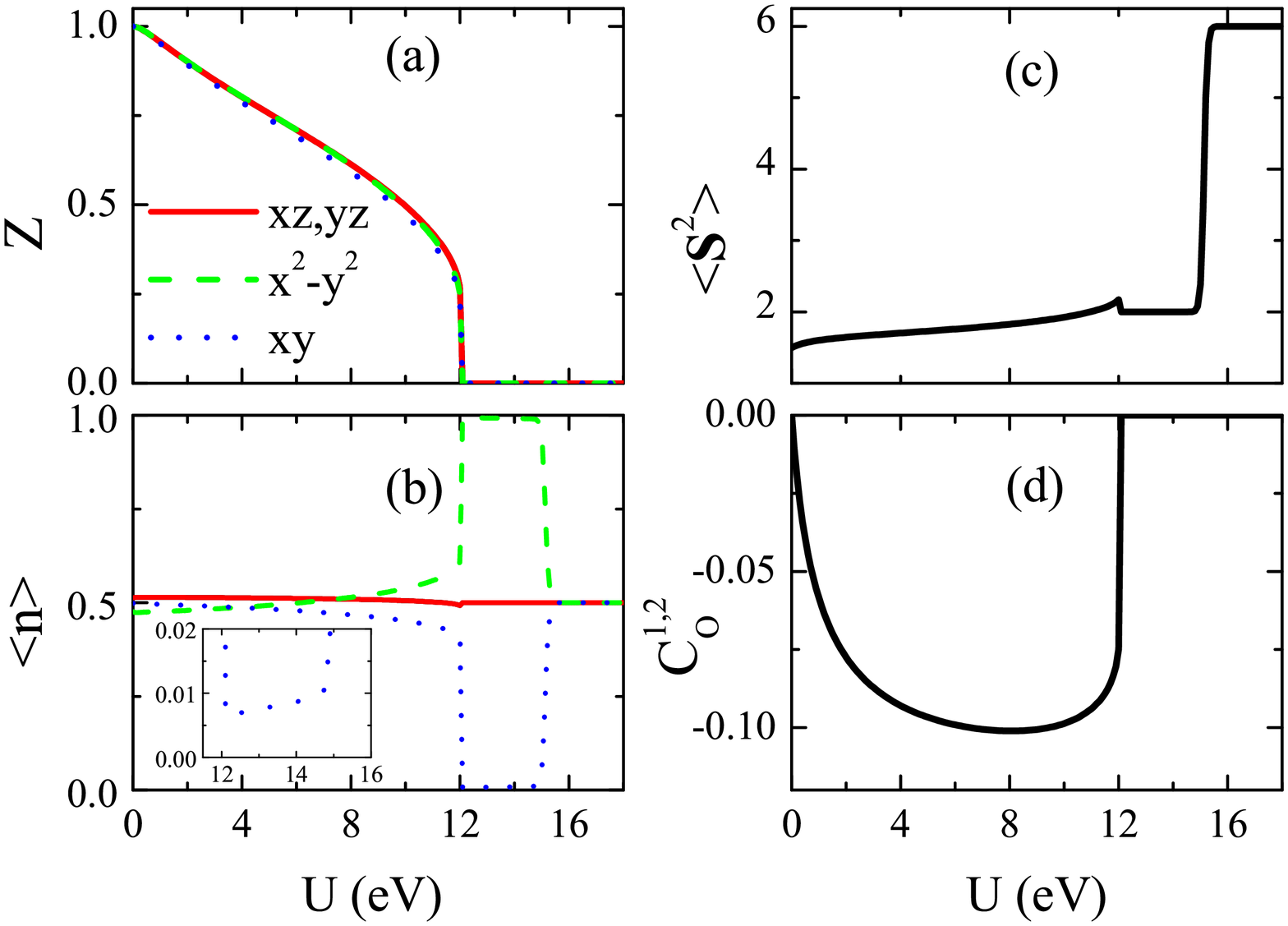}
\caption{(Color online) SSMF results for the four-orbital model at $J/U=0.007$.
}
\label{f.Fig8}
\end{figure}

\begin{figure}[h]
\includegraphics[
bbllx=20pt,bblly=20pt,bburx=710pt,bbury=590pt
clip
,width=80mm]
%[width=1.0\linewidth]
{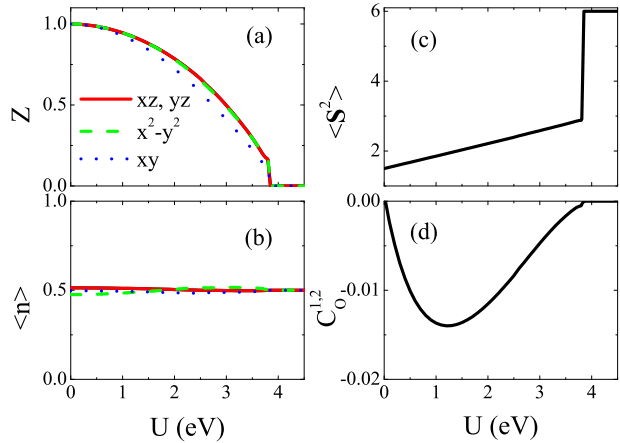}
\caption{(Color online) SSMF results for the four-orbital model at $J/U=1/4$.
}
\label{f.Fig9}
\end{figure}

\begin{figure}[h]
\includegraphics[
%bbllx=20pt,bblly=60pt,bburx=710pt,bbury=590pt
clip
,width=80mm]
%[width=1.0\linewidth]
{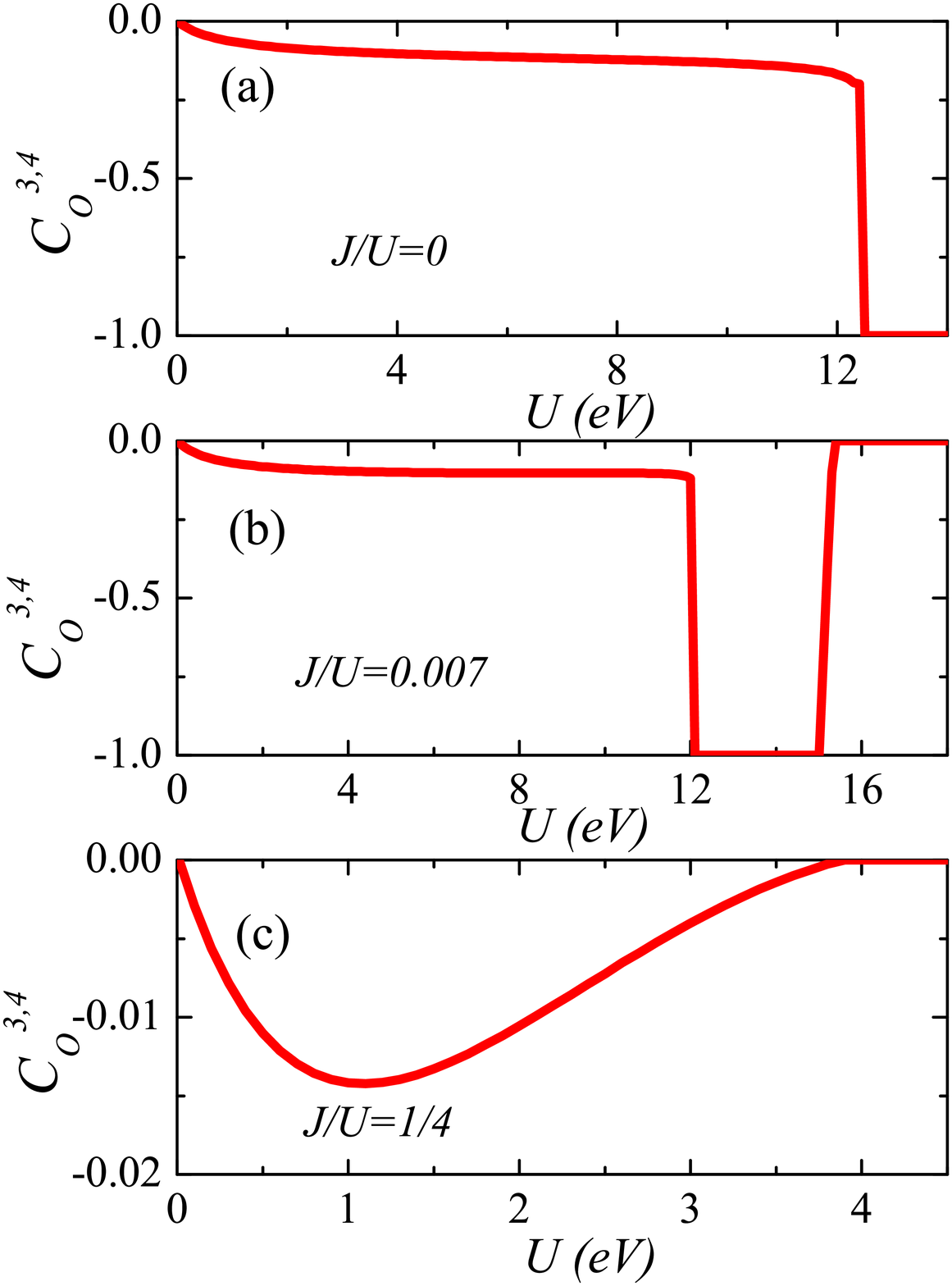}
\caption{(Color online) The evolution of inter-orbital correlation between the $xy$ and $x^2-y^2$ orbitals, $C_O^{3,4}$, in the four-orbital model at $J=0$, $J/U=0.007$, and $J/U=1/4$, respectively.
}
\label{f.Fig10}
\end{figure}

\begin{figure}[h]
\includegraphics[
%bbllx=20pt,bblly=60pt,bburx=710pt,bbury=590pt
clip
,width=80mm]
%[width=1.0\linewidth]
{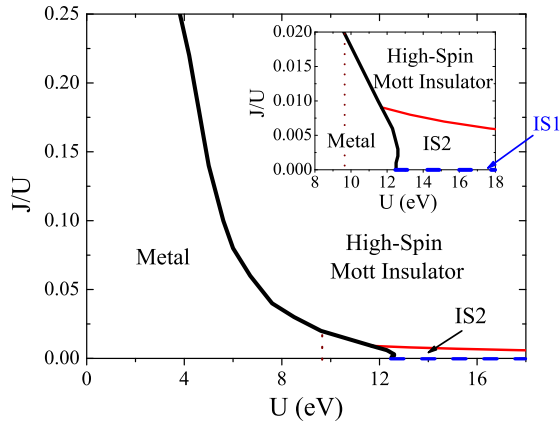}
\caption{(Color online) Phase diagram in the $J$-$U$ plane of the four-orbital model. The thick solid (black) line shows the phase boundary between the metallic and insulating states. The blue dashed line at $J=0$ refers to the intermediate-spin insulating state IS1, and the thin solid (red) line refers to the low-spin to high-spin transition between the IS2 state and the high-spin Mott state. The dotted line at $U\approx 10$ eV separates the metallic phase into two
regimes. To the left of this line, the Fermi surface topology is identical to the non-interacting case, while in the regime between this dotted line and the thick solid line, the topology of the Fermi surface changes (see text). Inset is a closer view of the lower-right corner of the phase diagram.}
\label{f.Fig11}
\end{figure}

In Fig.~\ref{f.Fig7} to Fig.~\ref{f.Fig11} we show the results from
SSMF calculation, in which the full Hund's coupling and the constraint
$U'=U-2J$ are taken into account. For both $J=0$ and $J>0$, an MIT is
observed. For $J=0$, $U_c=12.5eV$. A nonzero
$J$ may significantly reduce $U_c$. At $J/U=1/4$, $U_c$ is reduced to $3.82eV$. Interestingly, we find that the nature of the insulating state, and hence the MIT, is significantly affected by the competition between the Hund's coupling $J$ and the crystal field splitting $\Delta$ between the $xy$ and $x^2-y^2$ orbitals. For $J/U\gtrsim0.009$ the insulating state is a high-spin state with $S=2$ as illustrated in Fig.~\ref{f.Fig6}. This state is analogous to the high-spin Mott state discussed in the two-orbital model. One may directly check from Fig.~\ref{f.Fig9}(d) and Fig.~\ref{f.Fig10}(c) that the inter-orbital correlations are very small in the metallic state, and vanish in the insulating state, with a behavior similar to the two-orbital case at finite $J$.

The transitions at small and zero $J/U$ ratios are of special interest. As shown in Fig.~\ref{f.Fig7}(c), the insulating state at $J=0$ has an intermediate spin between $S=0$ and $S=1$.
We call this as IS1 state, which is shown in the phase diagram, {\it c.f.} Fig.~\ref{f.Fig11}. This transition
to the IS1 state can be understood as follow: at $J=0$, the $xy$ orbital is empty and the $x^2-y^2$ orbital is fully occupied in the insulating state due to the crystal field splitting. Hence at $U=U_c$ these two orbitals undergo a transition to a band insulator. On the other hand, the degenerate $xz$ and $yz$ orbitals are at half-filling, and a Mott transition takes place at the same $U_c$ value. Since at $J=0$ all the six $n=2$ configurations in $xz$ and $yz$ orbitals are degenerate, the Mott insulator is a mixture of $S=0$ and $S=1$ states. This gives the IS1 state an intermediate spin value. The transitions to the band insulator and Mott insulator are reflected in the behavior of the inter-orbital
correlation functions $C_O^{1,2}$ and $C_O^{3,4}$,
where $C_O^{\alpha,\beta} = \langle (n_{\alpha}-1)(n_{\beta}-1)\rangle$ for
$\alpha,\beta=1,2,3,4$. Here 1,2,3,4 denote $xz$, $yz$, $x^2-y^2$ and $xy$,
respectively. At $J=0$, $C_O^{3,4}$ jumps to $-1$ at $U_c$, indicating a
transition to a band insulator. But $C_O^{1,2}>-1$ even in the insulating phase,
 % indicates
signaling that the insulating state is a mixture of $S=1$ and $S=0$ Mott states, just as in the two-orbital model at $U=U'$.
In general, the Mott transition in the $xz$ and $yz$ orbitals may take place at a $U$ value different than $U_c$. But in this four-orbital model, $U_c$ for the transition to the band insulator in the $xy$ and $x^2-y^2$ orbitals is larger than the critical value for the Mott transition in two degenerate orbitals. Hence when the other two orbitals become band insulators at $U_c$ and thus are decoupled from the $xz$ and $yz$ orbitals, the $xz$ and $yz$ orbitals enter the Mott insulating state immediately via a first-order transition. We call this special transition at $U_c$ an orbitally selective MIT. This transition is different from the OSMT in that it takes place at a single $U_c$,
with different orbitals entering different insulating states.

We find that the orbitally selective MIT extends to
nonzero
$J$ values up to $J/U\approx 0.009$.
For nonzero $J$ in this range, the state on the insulator side has an intermediate spin value $S=1$.
We denote this state as the IS2 state.
Shown in Fig.~\ref{f.Fig8} is the
transition to the IS2 state in the case of
$J/U=0.007$, for which the
% is shown in Fig.~\ref{f.Fig8}. The
IS2 state is found to be stabilized for $12eV\lesssim U\lesssim15.2eV$. Similar to the IS1 state at $J=0$, the IS2 state is stabilized because different orbitals undergo transitions to different insulating states: an $S=1$ Mott insulator for the $xz$ and $yz$ orbitals, and an $S=0$ orbitally polarized insulator
for the $xy$ and $x^2-y^2$ orbitals. The orbitally polarized insulator can be understood by studying the atomic limit of an effective two-orbital model including $xy$ and $x^2-y^2$ orbitals.\cite{Werneretal07} In this effective model, the pair-hopping term couples the two states $|\uparrow\downarrow\rangle_{xy} |0\rangle_{x^2-y^2}$ and $|0\rangle_{xy} |\uparrow\downarrow\rangle_{x^2-y^2}
 $. When $\Delta>2\sqrt{2}J$, the ground state is a spin singlet $\cos\theta |0\rangle_{xy} |\uparrow\downarrow\rangle_{x^2-y^2} - \sin\theta |\uparrow\downarrow\rangle_{xy} |0\rangle_{x^2-y^2}$, where $\tan\theta = \sqrt{(\Delta/J)^2+1}-\Delta/J$. One may check that $n_{xy}=\sin^2\theta$,
 and $n_{x^2-y^2}=\cos^2\theta$. In general, each orbital is only partially occupied, and $n_{xy}\neq n_{x^2-y^2}$, which is clearly shown
 in Fig.~\ref{f.Fig8}(b) within $12eV<U<15.2eV$. Hence this state is different from either a band insulator or a Mott insulator, and is denoted as orbitally polarized insulating state. From Fig.~\ref{f.Fig10}, we see that the inter-orbital correlation in the orbitally polarized state behaves as same as in the band insulator. To understand this, note that there are two special limits in the orbitally polarized insulator. First, when $J=0$ and $\Delta>0$, $\sin\theta=0$
 and this state describes the band insulator with fully occupied $x^2-y^2$ orbital and empty $xy$ orbital. The other limit appears when $\Delta=0$
 but $J>0$. This leads to $\sin\theta=\cos\theta=1/\sqrt{2}$, and the state is identical to the low-spin orbital-Mott state discussed in Sec. IIIB.
 Thus the band insulator and the low-spin orbital-Mott insulator are adiabatically connected by the orbitally polarized insulating state.

Further increasing $J$ in the effective two-orbital model for $xy$ and $x^2-y^2$ orbitals eventually leads to a low-spin to high-spin transition.\cite{Werneretal07} The ground state manifold changes from the $S=0$ orbitally polarized state to the $S=1$ Mott state. In the four-orbital model, this corresponds to a first-order transition in the insulating states from the IS2 state to the high-spin Mott insulator, which takes place at $\Delta=2\sqrt{6}J$. Since in this study we assume $J$ is proportional to $U$, the low-spin to high-spin transition is accessible by increasing $U$. In Fig.~\ref{f.Fig8} we identify that it takes place at $U\approx15.2eV$ for $J/U=0.007$: the total spin jumps from $S=1$ to $S=2$, and the filling factors of $xy$ and $x^2-y^2$ orbitals
rapidly converge to half-filling.

In Fig.~\ref{f.Fig11} we show the $J$-$U$ phase diagram for the four-orbital model. For $J/U\gtrsim 0.009$ the MIT takes place between a paramagnetic metal and $S=2$ high-spin Mott insulator. In this regime, the phase diagram is similar to the one in the two-orbital model shown in Fig.~\ref{f.fig3}. The main difference from the phase diagram in Fig.~\ref{f.fig3} lies at $U\gg\Delta$ and $J\ll\Delta$. In this regime we find two intermediate-spin states: the IS1 state at $J=0$ and the IS2 state at finite $J$. The boundary between the IS2 and the $S=2$ Mott insulator, $J/U=\Delta/2\sqrt{6}U$, is determined by solving the Hamiltonian of the four-orbital model in the atomic limit. Given that $\Delta=0.52eV$ in this model, and assuming that $U_c$ for the MIT at $J/U\ll1$ stays the same value as $J=0$, we may estimate the tricritical point where the transition between the IS2 and the high-spin Mott state and the MIT meets. It is located at $J/U\approx0.0085$, which is consistent wit
 h the numerical result $J/U\approx0.009$.

It is very important to check how the Fermi surface in the metallic state evolves in the presence of interactions. In the noninteracting case, the Fermi surface of the four-orbital model consists of
two hole pockets centered at $(0,0)$ point in the one-iron Brillouin zone
and two electron pockets centered at $(\pi,0)$ and $(0,\pi)$ points,
respectively. We have checked that for a large portion of the metallic
regime in the phase diagram, i.e., to the left of the dotted line at $U\approx10$ eV in
Fig.~\ref{f.Fig11}, the Fermi surface of the interacting system has the
same topology as that for
% of the Fermi surface in
the non-interacting system. In this regime, even near the high-spin Mott insulator, there are only very tiny changes in the size and shape of the hole and electron pockets. This is
%consistent with
similar to the result
%in
of the two-orbital model, in which the Fermi surface is
always identical to the one in the non-interacting system. But in the four-orbital model, we find that the topology of the Fermi surface may
change when the system is close enough to the intermediate-spin states
(in the regime between the dotted line and the thicker solid line in
Fig.~\ref{f.Fig11}). In this regime, as
 %increasing $U$,
$U$ is increased,
two additional small electron pockets centered at $(\pi,0)$ and $(0,\pi)$ points
and one additional hole pocket centered at $(\pi,\pi)$ point may appear.
Such a change in the Fermi surface topology
primarily reflects
the difference in the electron occupancies of the different orbitals compared to the non-interacting case
(cf.~Figs.~\ref{f.Fig7}b, \ref{f.Fig8}b, \ref{f.Fig9}b),
which is a precursor to the orbitally-selective MIT occurring
in the small $J/U$ regime.
%the interplay
% %of
%between the
 % crystal field splitting and
 % % the
 % electron correlations.

MIT has also been discussed in spin density wave (SDW) calculations in a number of multi-orbital models for iron pnictides.\cite{Yuetal09,Daghofer09,Bascones10} An intermediate-spin insulating state is recently found in a five-orbital model within the SDW MF theory.\cite{Bascones10} Though a full study of the MIT in the magnetically ordered state using SSMF theory is beyond the scope of this paper, we find it quite interesting to compare the intermediate-spin states found in our study with those in Ref.~\onlinecite{Bascones10} in the atomic limit. In both works, the intermediate-spins states are found when $U/t$ is large but $J/U$ is small, indicating that these states all originate from the interplay of the crystal field splitting and Hund's coupling.
But there are some differences. First, in Ref.~\onlinecite{Bascones10}, a $S=0$ state violating the Hund's rule is stabilized at $J=0$. This state originates from the lifting of orbital degeneracy between the $xz$ and $yz$ orbitals by magnetic ordering. But In our case, the orbital degeneracy between the $xz$ and $yz$ orbitals stabilizes the IS1 state with a non-zero spin value at $J=0$. A second difference lies in the way treating the electron-electron interaction term $H_{int}$: we consider the full interaction in our SSMF calculation; while in Ref.~\onlinecite{Bascones10}, the spin-flip and the pair-hopping terms are neglected from the MF approximation.
This results in different insulating states: the IS2 state in our paper consists an orbitally polarized insulator in the $xy$ and $x^2-y^2$ orbitals, but the bands with $xy$ and $x^2-y^2$ orbital characters in the $S^z=1$ state in Ref.~\onlinecite{Bascones10} are simple band insulators.

An interesting question is whether an OSMT may take place when more than
two orbitals are included. Recently, a mechanism of the OSMT based on the lifting of orbital degeneracy is proposed.~\cite{Medici09} According to this mechanism, in a system with more than two orbitals, if the orbital degeneracy is partially lifted by a crystal field splitting, an OSMT may take place even when the bandwidths are equal. It is suggested
that an OSMT triggered by this mechanism may exist in iron pnictides.\cite{Medici10} From Fig.~\ref{f.Fig9} we find that for $J>0$, the bandwidth of
the $xy$ orbital gets stronger renormalization than the others. This effect
becomes more pronounced for larger $J$.
However, no OSMT is observed up to $J/U=1/2$. This is not too surprising. On the one hand, the above mechanism is proposed by assuming zero inter-orbital hopping. The four-orbital model discussed in this section contains non-zero
inter-orbital hoppings, which enhance the orbital correlations and favor
an one-stage Mott transition. On the other hand, according to the mechanism, an OSMT is easier to be realized if the
%lifted orbital
orbital whose degeneracy is lifted is at half-filling but the degenerate orbitals are away from half-filling,
so that the lifted orbital
is localized while the degenerate ones are still metallic. In the four-orbital model we consider,
the degenerate $xz$ and $yz$ orbitals are exactly at half filling, making the Mott insulator more stable.

\section{Conclusion}
In conclusion, we have studied the metal-insulator transitions in
several multi-orbital Hubbard models for the parent iron pnictides
using a slave-spin formulation.

In the two-orbital model, a transition to a
Mott insulator generally exists at half-filling. The Hund's coupling
reduces the critical coupling significantly.
We find that the nature of the Mott insulator depends on the ratio $U'/U$.
For $U'<U$, the insulator is a high-spin Mott state with zero double occupancy
but a spin triplet.
% whereas
For $U'>U$, by contrast, the insulator is an orbital-Mott state
with spin singlet and finite double occupancy. The low-spin orbital-Mott state
is unstable to a band insulator if the orbital degeneracy is lifted, and can be viewed as a special case of an orbitally polarized insulator.

The phase diagram for the metal-to-insulator transition in the more
realistic four-orbital model
%is rich.
contains additional features.
We find a transition to a high-spin $S=2$ Mott insulator when the Hund's coupling is strong enough ($J/U\gtrsim0.009$). This high-spin Mott insulator is an analogy to the $S=1$ Mott insulator in the two-orbital model.
At weak
%and
including
zero Hund's couplings ($J/U\lesssim0.009$) we find a transition to intermediate-spin insulating states. Such a transition is an orbitally selective metal-insulator transition, namely, different orbitals undergo transitions to different insulating states at a single critical value
$U_c$. At $J=0$ the transition leads to a band insulator for two
($xy$ and $x^2-y^2$) orbitals and a Mott insulator for the
other two ($xz$ and $yz$) orbitals. For
nonzero $J$ in this regime, the transition in the $xz$ and $yz$ orbitals
is to the Mott insulator, but
%due to the pair-hopping term,
an orbitally polarized insulator is stabilized in the $xy$ and $x^2-y^2$ orbitals on the insulating side
due to the pair-hopping term.
As $J$ is increased further, the intermediate-spin orbitally polarized insulator
%As further increasing $J$, the orbitally polarized insulator
%, and hence the  intermediate-spin state,
undergoes a low-spin to high-spin transition to a high-spin Mott insulator.

The existence of a Mott transition in the multi-orbital models with
an even number of electrons per Fe provides the theoretical basis
for the recent finding of a Mott insulator
in the iron oxychalcogenides with an expanded Fe lattice \cite{Zhu10}.
In addition, it strengthens the notion that the iron pnictides are
located in proximity to a Mott localization
transition.

%\ry{
{\it Note Added.} After this paper was first submitted for publication and
%appeared
posted
on the arXiv preprint listing,
several
% related
%works
studies
%have
also
%emphasized  on the similar issue appeared
discussed the relation between the band and orbital pictures in related
contexts \cite{Knolleetal11,Nicholson11},
and a number of works
%have
used the slave-spin or a related slave-roter method to investigate the metal-insulator transitions
in related multi-orbital models and systems \cite{Yu11,Zhou11,Ko11}.

%\begin{acknowledgments}
{\bf Acknowledgments}
We would like to thank L. Baksmaty and P. Goswami for useful discussions.
This work was supported by
the NSF Grant No. DMR-1006985,
 the Robert A. Welch Foundation
Grant No. C-1411, and the W. M. Keck Foundation.


\begin{thebibliography}{99}

\bibitem{Kamihara_FeAs}
Y. Kamihara,
T. Watanabe, H. Hirano, and H. Hosono,
J. Am. Chem. Soc. {\bf 130}, 3296 (2008).

\bibitem{Zhao_CPL08}
Z. A. Ren, W. Lu, J. Yang, W. Yi, X. L. Shen , Z. C. Li,
G. C. Che, X. L. Dong, L. L. Sun, F. Zhou, and Z. X. Zhao,
Chin. Phys. Lett. {\bf 25}, 2215 (2008).

\bibitem{delaCruz}
C. de la Cruz, Q. Huang, J. W. Lynn, J. Li, W. Ratcliff II,
J. L. Zarestky, H. A. Mook,  G. F. Chen, J. L. Luo, N. L. Wang,
and P. C. Dai, Nature {\bf 453}, 899 (2008).

\bibitem{Si-Abrahams-prl08} Q. Si and E. Abrahams, Phys. Rev. Lett.
{\bf 101}, 076401 (2008).

\bibitem{Si-Abrahams-Dai-Zhu} Q. Si, E. Abrahams, J. Dai, and J.-X. Zhu,
New J. Phys. {\bf 11}, 045001 (2009).

\bibitem{Haule-Kotliar-prl08} K. Haule, J. H. Shim, and G. Kotliar,
Phys. Rev. Lett. {\bf 100}, 226402 (2008).

\bibitem{Laad08}
%\qs{REF}
M. S. Laad, L. Craco, S. Leoni, and H. Rosner,
Phys. Rev. B {\bf 79}, 024515 (2009).

\bibitem{Daghofer08}
%\qs{REF}
M. Daghofer, A. Moreo, J. A. Riera, E. Arrigoni, D. J. Scalapino, and E. Dagotto, Phys. Rev. Lett.
{\bf 101}, 237004 (2008).

\bibitem{CFang08}
%\qs{REF}
C. Fang, H. Yao, W.-F. Tsai, J.-P. Hu, and S. A. Kivelson, Phys. Rev. B {\bf 77},
224509 (2008).

\bibitem{CXu08} C. Xu, M. Mueller, and S. Sachdev,
Phys. Rev. B {\bf 78}, 020501(R) (2008).

\bibitem{IshidaLiebsch} H. Ishida and A. Liebsch, Phys. Rev. B {\bf 81},
054513 (2010).

\bibitem{Leeetal09} H. Lee, Y.-Z. Zhang, H. O. Jeschke,
and R. Valent\'{\i},
%arXiv:0912.4024.
Phys. Rev. B {\bf 81},
220506(R) (2010).


\bibitem{Qazilbash}
%\qs{REF}
M. M. Qazilbash, J. J. Hamlin, R. E. Baumbach, L. Zhang, D. J. Singh, M. B. Maple, and D. N. Basov,
Nat. Phys. {\bf 5}, 647 (2009).

\bibitem{Hu08} W. Z. Hu,
J. Dong, G. Li, Z. Li, P. Zheng, G. F. Chen, J. L. Luo, and N. L. Wang,
Phys. Rev. Lett. {\bf 101}, 257005 (2008).

\bibitem{Si_natphys}
Q. Si, Nat. Phys. {\bf 5}, 629 (2009).

\bibitem{Yang08}
J. Yang
%\emph{et al.},
D. H\"{u}vonen, U. Nagel, T. R\~{o}\~{o}m, N. Ni,
P. C. Canfield, S. L. Bud'ko, J. P. Carbotte, and T. Timusk,
Phys. Rev. Lett. {\bf 102}, 187003 (2009).

\bibitem{Boris09}
A. V. Boris,
% \emph{et al.},
N. N. Kovaleva, S. S. A. Seo, J. S. Kim, P. Popovich,
Y. Matiks, R. K. Kremer, and B. Keimer,
Phys. Rev. Lett. {\bf 102}, 027001 (2009).

\bibitem{Yang09}
%\qs{REF}
W. L. Yang, A. P. Sorini, C-C. Chen, B. Moritz, W.-S. Lee, F. Vernay, P. Olalde-Velasco, J. D. Denlinger, B. Delley, J.-H. Chu, J. G. Analytis, I. R. Fisher, Z. A. Ren, J. Yang, W. Lu, Z. X. Zhao, J. van den Brink, Z. Hussain, Z.-X. Shen, and T. P. Devereaux,
Phys. Rev. B {\bf 80}, 014508 (2009).

\bibitem{Kutepov10}
%\qs{REF}
A. Kutepov, K. Haule, S. Y. Savrasov, and G. Kotliar,
% arXiv:1005.0885.
Phys. Rev. B {\bf 82},
045105 (2010).


\bibitem{Zhao} J. Zhao,
%\emph{et al.}, %
D. T. Adroja, D.-X. Yao, R. Bewley, S. Li, X. F. Wang,
G. Wu, X. H. Chen, J. Hu, and P. Dai,
Nat. Phys. {\bf 5}, 555 (2009).

\bibitem{Zhu10}
J.-X. Zhu, R. Yu, H. Wang, L. L. Zhao, M. D. Jones, J. Dai,
E. Abrahams, E. Morosan, M. Fang, and Q. Si,
Phys. Rev. Lett. {\bf 104}, 216405 (2010).

\bibitem{ImadaRMP98}
M. Imada, A. Fujimori, and Y. Tokura, Rev. Mod. Phys. {\bf 70}, 1039.

\bibitem{DagottoRMP94} E. Dagotto, Rev. Mod. Phys. {\bf 66}, 763 (1994).

\bibitem{GKKRRMP96} A. Georges, G. Kotliar, W. Krauth, and M. Rozenberg,
Rev. Mod. Phys. {\bf 68}, 13 (1996).

%\bibitem{Slater51} J. C. Slater, Phys. Rev. {\bf 82}, 538 (1951).
\bibitem{BrinkmanRice} W. F. Brinkman and T. M. Rice, Phys. Rev.
B {\bf 2}, 4302 (1970).

\bibitem{Ohashietal08} T. Ohashi, T. Momoi, H. Tsunetsugu, and N. Kawakami,
Phys. Rev. Lett. {\bf 100}, 076402 (2008).

\bibitem{Anisimovetal02}
%\qs{REF}
V. I. Anisimov, I. A. Nekrasov, D. E. Kondakov, T. M. Rice, and M. Sigrist,, Eur. Phys.
J. B {\bf 25}, 191 (2002).

\bibitem{Liebsch03}
A. Liebsch, Phys. Rev. Lett. {\bf 91}, 226401 (2003).

\bibitem{Kogaetal04} A. Koga, N. Kawakami, T. M. Rice, and M. Sigrist,
Phys. Rev. Lett. {\bf 92}, 216402 (2004).

\bibitem{Knechtetal05} C. Knecht, N. Bl\"{u}mer,
and P. G. J. van Dongen, Phys. Rev. B {\bf 72}, 081103(R) (2005).

\bibitem{AritaHeld05} R. Arita and K. Held, Phys. Rev. B {\bf 72},
201102(R) (2005).

\bibitem{Werneretal09} P. Werner, E. Gull, and A. J. Millis,
Phys. Rev. B {\bf 79}, 115119 (2009).

\bibitem{Leeetal0910} H. Lee, Y.-Z. Zhang, H. O. Jeschke, and
R. Valent\'{\i}, Phys. Rev. Lett. {\bf 104}, 026402 (2010).

\bibitem{Onoetal03} Y. Ono, M. Potthoff, and R. Bulla, Phys. Rev.
B {\bf 67}, 035119 (2003).

\bibitem{FlorensGeorges04} S. Florence and A. Georges,
Phys. Rev. B {\bf 70}, 035114 (2004).

\bibitem{Raghu}
%\qs{REF}
S. Raghu, X.-L. Qi, C.-X. Liu, D. J. Scalapino, and S.-C. Zhang, Phys. Rev. B {\bf 77},
220503(R) (2008).

\bibitem{deMedici05}
L. de'Medici, A. Georges, and S. Biermann,
Phys. Rev. B {\bf 72}, 205124 (2005).

\bibitem{deMedici10}
%\ry{REF}
S. R. Hassan and L. de'Medici, Phys. Rev. B {\bf 81}, 035106 (2010).

\bibitem{Lu} J. P. Lu, Phys. Rev. B {\bf 49}, 5687 (1994).

\bibitem{Rozenberg}
M. J. Rozenberg, Phys. Rev. B {\bf 55}, R4855 (1997).

\bibitem{HundRule} J. B\"{u}nemann, W. Weber, and F. Gebhard, Phys. Rev.
B {\bf 57}, 6896 (1998); K. Inaba, A. Koga, S.-I. Suga, and N. Kawakami,
Phys. Rev. B {\bf 72}, 085112 (2005); K. Inaba and A. Koga, Phys. Rev. B
{\bf 73}, 155106 (2006).

\bibitem{Castellani} C. Castellani, C. R. Natoli, and J. Ranninger,
Phys. Rev. B {\bf 18}, 4945 (1978); E. Dagotto, T. Hotta, and A. Moreo, Phys. Rep. {\bf 344}, 1 (2001).

\bibitem{KotliarRuckenstein} G. Kotliar and A. E. Ruckenstein,
Phys. Rev. Lett. {\bf 57}, 1362 (1986).


\bibitem{FlorensGeorges02} S. Florence and A. Georges,
Phys. Rev. B {\bf 66}, 165111 (2002).

\bibitem{SU2}
The preservation of the SU(2) symmetry can be seen by checking that
the pseudo-spin operators,
defined in terms of slave spin operators as
$\tau^+=S^+_\uparrow S^-_\downarrow$, $\tau^-=S^+_\downarrow S^-_\uparrow$,
and $\tau^z=(S^z_\uparrow-S^z_\downarrow)/2$ for each orbital,
satisfies the SU(2) spin commutation relation.

\bibitem{Dingetal08}
%\qs{REF}
H. Ding, K. Nakayama, P. Richard, S. Souma, T. Sato, T. Takahashi, M. Neupane, Y.-M. Xu, Z.-H. Pan, A.V. Federov, Z. Wang, X. Dai, Z. Fang, G.F. Chen, J.L. Luo, and N.L. Wang, arXiv:0812.0534

\bibitem{Hund2} Quantitatively, the perturbation theory works well only at $J\approx0$ since only at this point the MIT is continuous. At finite $J/U$, we find the MIT is discontinuous, similar to the result in a recent slave-boson study: F. Lechermann, A. Georges, G. Kotliar and O. Parcollet, Phys. Rev. B {\bf 76}, 155102 (2007). The discontinuous nature of MIT may cause the perturbation theory invalid.

\bibitem{Gunnarsson96} O. Gunnarsson, E. Koch, and R. M. Martin, Phys. Rev. B {\bf 54}, R11026 (1996); J. Han, M. Jarrell, and D. Cox, Phys. Rev. B {\bf 58}, R4199 (1998).

\bibitem{note1}
Note that $U'=U-2J$ is strictly satisfied for an isolated atom in free space, where the Coulomb interaction has the full rotational symmetry. In solids, the rotational symmetry is reduced. Taking into account the many-body effects between the electrons and ligand ions, this constraint may not be strictly satisfied. Specific to the two-orbital model with $xz$ and $yz$ orbitals, this constraint implies that
the interaction is invariant under a continuous rotation in the $xy$ plane (an $O(2)$ symmetry). However, this $O(2)$ symmetry is not necessarily satisfied in a solid, where only the discrete point group symmetry is respected. So the $U'=U-2J$ may not be always valid in a solid. For instance, as discussed in Ref.~\onlinecite{Yanagi10}, it is broken due to the electron-phonon interaction in pnictides. Although Ref.~\onlinecite{Yanagi10} discussed a generalized $16$-band model including both Fe $3d$ and As $4p$ orbitals, the electron-phonon interaction does not lift the orbital degeneracy. Hence by breaking the constraint $U'=U-2J$, our two-orbital model may just serve as a minimal model with effective $U$, $U'$ and $J$ taking account for the electron-phonon interaction.

\bibitem{Yanagi10}
Y. Yanagi, Y. Yamakawa, and Y. \={O}no, Phys. Rev. B {\bf 81}, 054518 (2010).

\bibitem{note}
Relaxing the constraint $U'=U-2J$ corresponds to picking up a specified set of orbital basis and making the couplings basis dependent since the continuous rotational symmetry in the orbital space is broken. This leads to the two spin-singlet states $1/\sqrt{2}(|\uparrow\downarrow;0\rangle - |0;\uparrow\downarrow\rangle)$ and $1/\sqrt{2}(|\uparrow;\downarrow\rangle - |\downarrow;\uparrow\rangle)$, with energies $U-J$ and $U'+J$, respectively, are no longer degenerate. But relaxing the constraint will not lift the orbital degeneracy of the $xz$ and $yz$ orbitals. Therefore our discussion on the Mott transition in Sec. IIIA is still valid.

\bibitem{note2}
Performing the perturbative treatment as in Sec. IIIA, we find for $U'>U$,
$U_c=4|\bar{\epsilon}|/(2U'/U-1-J/U)$. $U_c$ decreases
monotonically when increasing $U'/U$ from 1.

\bibitem{Graser} S. Graser, T. A. Maier, P. J. Hirschfeld, and D. J. Scalapino, New J. of Phys. {\bf 11}, 025016 (2009).

\bibitem{Kuroki}
%\qs{REF}
K. Kuroki, S. Onari, R. Arita, H. Usui, Y. Tanaka, H. Kontani, and H. Aoki,
Phys. Rev. Lett. {\bf 101}, 087004 (2008).
%\bibitem{Graser10} S. Graser \emph{et al.}, arXiv:1003.0133.

\bibitem{Werneretal07}
%\ry{REF}
P. Werner and A. J. Millis, Phys. Rev. Lett. {\bf 99}, 126405 (2007).

\bibitem{Yuetal09}
%\ry{REF}
R. Yu, K. T. Trinh, A. Moreo, M. Daghofer, J. A. Riera, S. Haas, and E. Dagotto, Phys. Rev. B 79, 104510 (2009).

\bibitem{Daghofer09}
%\ry{REF}
M. Daghofer, A. Nicholson, A. Moreo, and E. Dagotto, Phys. Rev. B {\bf 80}, 104507 (2009).

\bibitem{Bascones10}
%\ry{REF}
E. Bascones, M. J. Calderon, and B. Valenzuela, Phys. Rev. Lett. {\bf 104}, 227201 (2010).


%\bibitem{Ranetal09}
%\ry{REF}
%Y. Ran, F. Wang, H. Zhai, A. Vishwanath, and D.-H. Lee, Phys. Rev. B {\bf 79}, 014505 (2009).

\bibitem{Medici09} L. de'Medici, S. R. Hassan, M. Capone, and X. Dai,
Phys. Rev. Lett. {\bf 102}, 126401 (2009).

\bibitem{Medici10} L. de'Medici, S. R. Hassan, and M. Capone,
J. Supercond. Nov. Mat. {\bf 22}, 535 (2009).

\bibitem{Knolleetal11} J. Knolle, I. Eremin, and R. Moessner, Phys. Rev. B {\bf 83}, 224503 (2011).

\bibitem{Nicholson11} A. Nicholson, Q. Luo, W. Ge, J. Riera, M. Daghofer, G. B. Martins, A. Moreo, and E. Dagotto, Phys. Rev. B {\bf 84}, 094519 (2011).

\bibitem{Yu11}
R. Yu, J.-X. Zhu, and Q. Si,
Phys. Rev. Lett. {\bf 106}, 186401 (2011).

\bibitem{Zhou11}
Y. Zhou, D.-H. Xu, F.-C. Zhang, and W.-Q. Chen,
EPL {\bf 95}, 17003 (2011).

\bibitem{Ko11}
W.-H. Ko and P. A. Lee,
Phys. Rev. B {\bf 83}, 134515 (2011).

\end{thebibliography}
\end{document}